\documentclass{article}
\usepackage{geometry}[margin=1in]
\usepackage{graphicx} 
\usepackage{amsmath}
\usepackage{amssymb}
\usepackage{url}
\usepackage{mathrsfs}
\usepackage{authblk}
\usepackage{braket}
\usepackage{natbib}
\usepackage{enumitem}
\usepackage{tikz}

\usepackage{hyperref}
\usetikzlibrary{arrows.meta}
\title{Feasibilism, Explication, and the Cobham-Edmonds Thesis}
\author{Abrahim Ladha\thanks{Georgia Institute of Technology. Email: \href{mailto:abrahimladha@gatech.edu}{abrahimladha@gatech.edu}} \quad
Yiran Luo\thanks{Georgia Institute of Technology. Email: \href{mailto:yluo432@gatech.edu}{yluo432@gatech.edu}} \quad
Alan Tian\thanks{Carnegie Mellon University. Email: \href{mailto:alantian@andrew.cmu.edu}{alantian@andrew.cmu.edu}} 
}
\renewcommand{\P}{\mathsf{P}}
\newcommand{\NP}{\mathsf{NP}}
\newcommand{\EXP}{\mathsf{EXP}}
\newcommand{\TIME}{\mathsf{TIME}}

\newcommand{\PSPACE}{\mathsf{PSPACE}}
\newcommand{\BPP}{\mathsf{BPP}}
\newcommand{\BQP}{\mathsf{BQP}}
\newcommand{\FP}{\mathsf{FP}}
\newcommand{\PV}{\mathsf{PV}}
\newcommand{\APC}{\mathsf{APC}}
\newcommand{\SIZE}{\mathsf{SIZE}}

\begin{document}

\maketitle

\begin{abstract}
While the Church-Turing thesis asserts that effective calculability explicates to sets decidable by a Turing machine, the Cobham-Edmonds thesis asserts that feasible computation explicates to the complexity class $\P$, those decidable by a polynomial-time bounded Turing machine. The Church-Turing thesis has been placed under rigorous scrutiny and has several convincing arguments in its favor, but the Cobham-Edmonds thesis has not undergone a similar examination. Many of the arguments in its favor simply suggest that $\P$ is a useful assumption, rather than a necessary target. This paper presents analogous arguments in favor of the Cobham-Edmonds thesis.
\end{abstract}

\tableofcontents

\section{Background}
Structural complexity classes are bounded versions of classes in computability theory. While computable languages are those which have a Turing machine which halts on all inputs eventually, languages in $\P$ are those which have a Turing machine which halts within a number of steps bounded by a polynomial in the size of the input. While the computably enumerable languages are those which can be witnessed, $\NP$ is the class of languages which can be witnessed by a verifier with a polynomial-time bound.

The classes in computability theory are of great philosophical interest. By the Church-Turing thesis, anything computable, in the prescientific, intuitive sense, must be computable by a Turing machine, in the formal sense. Although the strict separation of the computable from the computably enumerable languages is a purely formal result, it provides great explanatory power into the informal concept of computation. By analogy, there should be philosophical interest in complexity classes as well. $\P$ is often characterized as problems which are \textit{feasibly computable}, and $\NP$ is often characterized as problems which are \textit{feasibly verifiable}. A popular phrasing of the unsolved $\P$ vs $\NP$ problem is then ``\textit{are all feasibly verifiable problems feasibly computable?}'' If this phrasing is to make any sense, it must be the case that the formal class $\P$ rigorously captures the intuitive notion of feasible computation. This work provides a detailed analysis of this problem.

\subsection{Historical and Philosophical Remarks}

The problem of feasible computation can be viewed as a continuation of the foundational debates of the twentieth century. Cantor's set theory, together with the subsequent formalization and axiomatization of mathematics, made the status of infinite totalities a central issue in the philosophy of mathematics. On one side stood the classical set-theoretic tradition, associated with Cantor and later axiomatized by Zermelo and Fraenkel, which permitted actual infinities and non-constructive methods. On the other side emerged a family of constructivist reactions, each attempting to restrict mathematical existence to what can be constructed, exhibited, or justified by more transparent means. Hao Wang classifies these foundational standpoints into five broad domains: anthropologism, finitism, intuitionism, predicativism, and platonism \citep{wang1958eighty,Wang1975-WANFMT-4}. Feasibilism belongs naturally to this landscape, since it asks not merely what exists or what is computable, but what can be carried out within admissible limits.

Throughout these debates, infinity remained one of the central axes of disagreement. Roughly speaking, constructivist positions tend to resist the unrestricted use of actual infinity, while classical mathematics permits reasoning over completed infinite totalities. Yet the most restrictive alternatives face their own difficulties. Strict finitism, for instance, attempts to admit only numbers and constructions that are feasible in practice; but this immediately raises the problem of drawing a non-arbitrary boundary between feasible and infeasible magnitudes. Yessenin-Volpin's ultrafinitist program made this issue explicit through the idea of feasible numbers, and Dummett's soritical objection showed how unstable such a predicate can become: if $n$ is feasible, then it seems that $n+1$ should also be feasible, although not every number can be feasible \citep{dummett1975wang}. The resulting tension is clear: if actual infinity is too permissive, while strict finitism is too restrictive and vague, then the locus of feasible mathematical practice must lie somewhere between unrestricted classical existence and merely finite surveyability \citep{gorbow2026feasible}.

Computability theory initially seemed to offer a natural way to sharpen this middle ground. The Church--Turing thesis identifies effective calculability with Turing computability, thereby giving a precise formal counterpart to the informal notion of an algorithmic procedure. Later in proof theory and type theory, the Curry--Howard correspondence further reinforces the connection between construction and computation: intuitionistic proofs can be interpreted as programs, and propositions as types. Nevertheless, computability alone is still too weak as an account of feasibility. A computable function may require resources far beyond any practical or physical bound; even primitive recursive functions, though total and effectively specified, include functions whose growth rates are far too large to serve as plausible models of feasible calculation. Bernays already observed that sufficiently large numerical expressions, such as $67^{(257^{729})}$, cease to have any concrete meaning from the standpoint of actual human computation \citep{bernays1935platonisme}. Computable in principle, therefore, does not by itself imply computable in practice.

This gap between abstract computability and feasible computation motivated a more restrictive line of inquiry. Wang's category of ``anthropologism'' explicitly concerns concepts tied to humanly surveyable or executable mathematical activity \citep{wang1958eighty}. In a more technical direction, Cobham's characterization of polynomial-time computable functions by limited recursion on notation provided a machine-independent way of isolating a robust class of feasibly computable functions \citep{Cobham1965-COBTIC}. This development led naturally to what is now called the Cobham--Edmonds thesis: the proposal that tractable or feasible computation is to be identified, at least extensionally, with polynomial-time computation.

The Cobham--Edmonds thesis gives feasibility a powerful mathematical explicatum, namely the class $\mathsf{P}$ for decision problems and $\mathsf{FP}$ for function computation. Its success lies in the stability of polynomial time across ``reasonable'' models of computation and in its central role in complexity theory. But its philosophical status is less straightforward. The thesis does not follow directly from the Church--Turing thesis; it adds a further claim about which resource bounds should count as feasible. Nor is it obvious that all and only polynomial-time procedures match the intuitive notion of feasibility. These worries have been noted in the literature \citep{gurevich1993feasible,dean2019computational}, yet much of bounded arithmetic proceeds by taking polynomial time as the operative formalization of feasible reasoning. Cook's theory $\PV$ and Buss's $S^1_2$, for example, are built around polynomial-time functions and thereby formalize feasibility through the Cobham--Edmonds lens. The remaining philosophical question is whether this identification is merely technically fruitful, or whether it gives the right explication of feasibility itself.

\subsection{Explication and Feasibilism}
Explication, in Carnap's sense, is ``\textit{the transformation of an inexact, prescientific concept, the explicandum, into a new exact concept, the explicatum}'' \citep{Carnap1950-CARLFO}. The explicandum is a concept whose ordinary use is often vague and only informally understood in natural language, whereas the explicatum is a formally specified concept which preserves the ordinary use of the explicandum while serving a theoretical purpose with greater exactness, fruitfulness, and simplicity. 

Carnap's own treatment of probability provides a famous example. The ordinary term ``probability'' does not correspond to a single precise concept. Rather, Carnap distinguishes at least two pre-theoretic uses: probability$_1$, understood as a degree of confirmation of a hypothesis by evidence, and probability$_2$, understood as relative frequency in the long run. The task of explication is therefore not to discover the one true meaning of the word ``probability'', but to clarify different uses of the term and provide precise formal counterparts for them.

But what precisely is a feasible procedure? The Church--Turing thesis provides a rare and unusually successful case of explication: an intuitive notion, that of an effective procedure carried out by a human computor according to fixed rules, is replaced by an exact mathematical notion, and the force of this replacement is strengthened by the convergence of several independent models of computation \citep{Turing1936-TUROCN,quinon2021church}. In this respect, Turing computability may be regarded as a paradigmatic explication of effective procedure. Feasibility, and also most other intuitive concepts, however, appears less amenable to such a single and canonical treatment. What counts as feasible varies with the theoretical context: in the field of property testing, for example, specialists treat even reading the entire input as infeasible, whereas in ordinary algorithmic contexts this is usually taken for granted. In contrast to this, many would consider computing the $n$th Fibonacci number to be feasible---the standard pen-and-paper algorithm being just iterated addition---but, when $n$ is given in binary as is standard, $F_n$ has value about $2^{0.69n}$ and hence has size $\Theta(n)$ bits, so merely writing down $F_n$ takes $2^{\Omega(m)}$ steps in the input length $m=\Theta(\log n)$. This suggests that feasibility is closer to the pluralistic side of Carnapian explication: formalization does not merely recover a preexisting, objective essence, but often selects and sharpens one use of an ambiguous intuitive concept at the expense of others, and therefore is subject to the decision of the explicator\cite{Carnap1950-CARLFO,dutilhNovaesReck2017carnapian}. Ideally, one would like an explication of feasibility as compelling as Turing's explication of effective procedure; but our investigation will demonstrate some difficulty in achieving this.

Although the notion of feasible computation is vague, any adequate explication should at least preserve two constraints. First, feasibility must be a subclass of computability: whatever is feasibly computable must be computable in the ordinary Turing sense, since feasibility adds resource constraints rather than removes the basic limits of computation. Second, this inclusion must be proper, otherwise the value of discussion is trivialized.

\subsection{Computational Complexity}
For functions $f,g:\mathbb{N}\rightarrow \mathbb{R}^+$, we say $f(n) \text{ is } O(g(n))$ if there exist constants $c,n_0 \in \mathbb{N}$ such that for all $n \geq n_0$ it is true that $f(n) \leq c \cdot g(n)$. Define $\TIME[f(n)]$ to be the class of languages which have multi-tape Turing machines to decide them within $O(f(n))$ steps on inputs of size $n$. Define $\P = \cup_{k = 1}^\infty \TIME[n^k]$. Similarly, let $\FP$ be the class of functions such that $f \in \FP$ if and only if $f$ is computable by a multi-tape Turing machine with a polynomial-time bound. For simplicity, we mostly focus on $\P$, but one can prove that a language is in $\P$ if and only if its characteristic function is in $\FP$. For more than basic notation in computational complexity, the reader is referred to the standard textbooks \cite{papadimitriou1994computational,sipser1996introduction,goldreich2008computational,arora2009computational,du2011theory}. For an introduction to $\PV$ for nonspecialists, see the work of Li \cite{li2025introduction}. For more on bounded arithmetic and complexity theory, see the book by Kraj{\'i}\v{c}ek \cite{krajicek1995bounded}. An article by Dean serves as an excellent introduction into computational complexity, written for philosophers \cite{dean2019computational}.

\subsection{Cobham and Edmonds}
We discuss the papers of Cobham and Edmonds.
The main contribution of Alan Cobham's seminal paper, \textit{The Intrinsic Computational Difficulty of Functions} \cite{Cobham1965-COBTIC} is the definition, and use of the definition of the class $\P$. He first observes that some functions high in the Grzegorczyk hierarchy appear hard to compute. After defining time and space complexity measures, he next observes that several basic arithmetic operations, including addition and multiplication, appear to have polynomial-time algorithms, thus the class $\P$ is a natural one to consider. He explains that the class $\P$ is invariant to many underlying changes to the variant of Turing machine under consideration, giving confidence that it must characterize intuitively feasible computation.

Jack Edmonds' paper, \textit{Paths, Trees, and Flowers} \cite{edmonds1965paths} appeared the same year as Cobham's. He gave an algorithm for finding a maximum matching in a graph (which we shall detail later as a useful example). Our main interest in Edmonds' work is a section he titles ``digression.'' He observes that although the theory of perfect matchings had been well developed by graph theorists---for example, Tutte's characterization of graphs with perfect matchings \cite{tutte1947factorization}---their methods left no procedure to actually compute a matching other than brute force search. He defines what he means by an efficient algorithm, one whose complexity ``only grows algebraically'' with the size of the input. This is to justify to the reader why his polynomial-time algorithm for finding a maximum matching is meaningful.

We have highlighted their differences, but there are many more similarities of the exposition of the problem in both papers. Both Cobham and Edmonds mention it is a difficult problem to consider exactly what constitutes an atomic step of an algorithm, a difficulty if a complexity measure is supposed to count a number of steps. They both explicitly do not commit to feasibility being associated with time, space, or a combination of both. They both operate under asymptotic analysis, observing it appears more intrinsic to the complexity of a problem. They both conjecture the existence of infeasible problems. Cobham mentions primality testing and Edmonds mentions graph isomorphism. Furthermore, they both mention that the problem of study bears conceptual resemblance to the Church-Turing thesis. The suggestion that $\P$ characterizes feasible computation has come to be denoted as the \textit{Cobham-Edmonds thesis}. Importantly, they both do not say that $\P$ must necessarily be the explicatum of feasible computation. They both simply suggest that there are many compelling reasons to accept $\P$. 

\section{\texorpdfstring{$\P$}{P} as a Plausible Hypothesis}
\label{prior}
The preceding section introduced the Cobham--Edmonds thesis as the proposal that feasible computation is to be explicated onto $\P$. The present section examines the general considerations that make this proposal attractive. Much of early complexity theory studied resource use comparatively: given two problems, machines, or algorithms, one could ask whether one requires more resource than the other \cite{hartmanis1965computational,trakhtenbrot2008survey}. Explication, however, requires more than such comparisons. Once computability has been accepted as too broad a notion of feasibility, one must still explain where, within the computable, the boundary between feasible and infeasible computation should be drawn, and in terms of what resources.

The arguments considered in this section are general in character. They do not derive $\P$ from the intuitive notion of feasibility, nor do they show that feasible computation \emph{must} be polynomial-time computation. Rather, they explain why $\P$ is a plausible and stable candidate for the role: it avoids the pathologies of non-uniform models, is robust under changes of reasonable machine model, is closed under ordinary algorithmic combinations, and treats polynomial-time solvability as a meaningful threshold in mathematical practice. These considerations support the polynomial-time explicatum without making it conclusive.

\subsection{Uniformity}
We argue that the explication of feasible computation must map to a uniform class, one in which a single device decides a given language for all inputs. A language is decidable by a uniform model of computation if there exists a single procedure deciding inputs of all lengths. Non-uniform models of computation exist, and include boolean and arithmetic circuits, neural networks, decision trees, and advice machines. For example, a circuit family $\mathcal{C} = \{C_0,C_1,C_2,...\}$ is an infinite set of circuits where $C_n$ has $n$ inputs and one output. We say a circuit family decides a language $L$ if $$w \in L \iff C_{|w|}(w) = 1$$ A non-uniform model of computation is an infinite family of devices. Each device is responsible only for inputs of a certain length. We need not infer anything from a 32-bit adder circuit failing to add 40-bit numbers. Let $\SIZE[f(n)]$ be the class of languages which have circuit families such that the number of gates of $C_n$ is $O(f(n))$. A fashionable class for problems with feasibly sized circuits is $\SIZE[poly] = \cup_{k=0}^\infty \SIZE[n^k]$.

We first prove that every class decidable by a non-uniform model of computation must be uncountable. Let $L \subseteq 1^*$ be a unary language. Consider a family of devices $\mathcal{D} = \{D_0,D_1,...\}$ such that if $1^n \in L$, then $D_n$ accepts only $1^n$, and if $1^n \not \in L$ then $D_n$ accepts no string. Certainly then $\mathcal{D}$ is a family of devices to decide $L$, and every such unary language has a fixed family of devices. Since $1^*$ is countably infinite, its powerset $\mathcal{P}(1^*)$ is uncountable by Cantor's theorem. Therefore the number of languages decidable by families of devices of this kind must be uncountable. 

Why do we detail this? It is uncontroversial that we require problems which are feasibly computable to be computable at all. The Turing-computable languages are countably infinite, and thus, cannot contain any uncountable subset. We therefore must conclude that the concept we explicate feasibility onto cannot be computed by a non-uniform model of computation. 

This issue with non-uniform models is a technical one, because it is not required that the function $f(1^n) = \braket{D_n}$ be computable. Technically, there are undecidable languages with linear size circuits. The study of circuit complexity is to only study the size, and not the complexity of synthesis of the circuit, so these properties are unfortunate accessories to the definitions. If one were to require that the circuits themselves be the output of some fixed polynomial-time algorithm, then this uniform version of $\SIZE[poly]$ is simply equivalent to the uniform class $\P$.  

\subsection{Stratospheric but within \texorpdfstring{$\P$}{P}}
A commonplace criticism of $\P$ representing feasible computation is that it contains languages decidable by algorithms which are technically polynomial time, but have very large degree. These appear intuitively infeasible. By the time hierarchy theorem, there must exist languages which require $\Omega(n^k)$ time for each $k$, and cannot be solved asymptotically faster. We argue these are of no concern. The time hierarchy theorem is proven via diagonalization, and the languages of arbitrarily high complexity have no use other than to be constructed to not have low complexity. An overwhelming majority of problems in $\P$ of actual interest appear to have a polynomial of low-degree. We detail three examples to address these concerns. Note that our argument is empirical and not theoretical. 

\begin{itemize}
\item Consider the problem of linear programming. Let $A$ be an $m\times n$ matrix, $b$ be a vector of length $m$, and $c$ be a vector of length $n$, all with rational elements. The problem of linear programming is to compute an $n$ length vector $x$ to maximize $c^Tx$ subject to $Ax \leq b$. Introduced in 1947, Dantzig's simplex method is the first known algorithm to solve this problem, but it does so in $O(2^n)$ steps in the worst case. In 1987, Khachiyan introduces the ellipsoid method, solving the problem in $O(n^4L)$ steps, where $L$ is the number of bits it takes to encode $A,b,c$ \cite{khachiyan1980polynomial}. It is theoretically significant that linear programming is solvable in polynomial time at all, but the degree of the polynomial was too large for practical application. In 1984, Karmarkar \cite{karmarkar1984new} invented the interior point method, solving the problem in $O(n^{3.5}L)$. After that, the complexity of the problem quickly fell. Vaidya was able to improve the complexity first to $O(n^3L)$ \cite{vaidya1987algorithm}, then to $O(n^{2.5}L)$ \cite{63499} using different ideas. The current state of the art has tied the problem to matrix multiplication with a time complexity of $O(n^{2.38})$\cite{jiang2020fasterdynamicmatrixinverse}, and improving this is an active area of research.

\item Given a graph $G = (V,E)$, a maximum matching is a set of edges $M \subseteq E$ of largest cardinality such that no two edges in $M$ share a vertex. In 1965, Edmonds published the blossom algorithm \cite{edmonds1965paths}, finding a maximum matching in $O(|V|^2|E|)$. Although not particularly fast, the best prior work was trivial brute force search in time $O(2^{|E|})$. The methods used in his algorithm have been described as a ``breakthrough in polyhedral combinatorics'' \cite{schrijver2003combinatorial}. 
After Edmonds' algorithm, the problem was improved to $O(|V||E|)$ \cite{gabow1976efficient} and eventually $O(\sqrt{|V|}|E|)$ \cite{DBLP:conf/focs/MicaliV80,vazirani2020proof}. No improvement has been found in four decades. 

\item Consider the problem of factoring polynomials with integer coefficients over the rationals. Early work was done by the Babylonians and Greeks for finding the roots of quadratics and some cubics. Early general methods were discovered by Bernoulli in 1702, von Schubert in 1793, and later rediscovered by Kronecker in 1882 \cite{RHM_2001__7_1_67_0}. To determine if a polynomial of degree $n$ has a factor of degree at most $m < n$, the complexity of their methods is roughly on the order of $O(2^m)$. Other exponential time methods were developed, such as the Zassenhaus method, but in 1982, there was a significant breakthrough. The Lenstra-Lenstra-Lov\'asz algorithm---which computes a reduced basis of a lattice in polynomial time---could be used in a special case for polynomial factorization \cite{Lenstra1982}. Its worst case analysis was on the order of $O(n^8)$ before being improved to $O(n^6)$ \cite{10.1007/3-540-13345-3_40}. Although initially still infeasible in practice, improvements have been made since. Some lattice reduction algorithms have specialized to assume the lattice is only of a certain form, which is still sufficient to solve polynomial factorization problems. The state of the art appears to be $O(n^3)$, but there are further non-asymptotic improvements \cite{van2012gradual}.

\end{itemize}
The moral of these problems is the same. Many problems lie in $\EXP = \cup_{k=0}^\infty \TIME[2^{n^k}]$ for no creative reason; they are simply na\"ively solvable by brute force search. Empirically, it appears to be the case that it takes a significant regime change to bring the problem within $\P$. Perhaps new mathematics are developed, a heuristic is proven correct, or an invariant of the problem is found. Once this occurs, methods are refined far easier than they are discovered. Problems solvable by algorithms with high-degree polynomial run time do not appear to stay high-degree. Algorithms only get faster, and there are almost no examples of problems whose fastest algorithm is a high-degree polynomial. What we describe is, of course, not a formal notion at all, but a social phenomenon. Improvements need not be at the asymptotic level either. Hidden constants improve, special cases are branched upon and handled by several algorithms, and optimizations are made to be more hardware friendly. 

\subsection{Closure}
If high-degree polynomial-time appears sparse, why consider it at all still? Why not simply explicate to strict bounds such as $\TIME[n]$ or $\TIME[n \log n]$ or $\TIME[n^2]$ rather than $\P$? The class $\P$ has many desirable closure properties that the former classes do not. Feasible computation, in the intuitive sense, is closed under many elementary combinations. Suppose $f,g$ are intuitively feasibly computable functions. Consider the operation of \textit{sequential execution}, where one executes one algorithm, followed by another. Sequential execution can be used to compute the composition of functions $g \circ f$. If $f,g$ have time bounds $T_f(n)$ and $T_g(n)$, then the sequential execution of $f$ then $g$ will take time on the order of $T_f(n) + T_g(n)$. Conveniently, polynomials are closed under addition, since $O(n^k) + O(n^l) = O(n^{\max(k,l)})$. There are elementary operations which intuitively feasible functions are closed under, such that the time complexity of the combination is a polynomial, assuming the components are. Consider the operation: for each step of $f$, run algorithm $g$. Such a combination would have run time $T_f(n) \cdot T_g(n)$, and conveniently, polynomials are closed under multiplication. $\P$ is closed under many forms of reduction since polynomials themselves are closed under composition. One can easily prove that $\FP$ is the smallest class of functions which contains the quadratic time computable functions and are closed under these operations \cite{du2011theory}. 

\subsection{The Extended Church Turing Thesis}
\label{ectt}
One common presentation of the Church-Turing Thesis (not advocated this way by either Church or Turing) is that the class of computably enumerable languages are independent of any specific formalism. It is styled as a falsifiable scientific hypothesis \cite{sipser1996introduction,lewis1998elements,hopcroft2001introduction}. Suppose that one were to propose a new model of computation with the hope of explicating the informal notion of algorithm to this model. It is likely that the model of computation is simulatable by those models which are already known, such as the $\lambda$-calculus or the Turing machine. The fact that all these Turing-complete models of computation have proofs of simulation for each other is the evidence commonly shown in favor of the Church-Turing thesis. Proofs in computability theory may proceed without difficult programming arguments in specific formalisms by taking the Church-Turing thesis as a premise \cite{Post1944}.

During the historical development of computational complexity, models of computation which had been assumed to be equivalent with respect to computability, were now different with respect to complexity. For example, a multi-tape Turing machine may decide if a word is a palindrome in $O(n)$ steps, but a single-tape Turing machine must take $\Omega(n^2)$ steps \cite{hennie1965one}.
The extended Church-Turing thesis (also called the feasibility thesis \cite{cook2000p}, the computational complexity-theoretic Church–Turing thesis \cite{bernstein1993quantum}, the invariance thesis \cite{etde_5635755}, or the strong form of the Church-Turing thesis \cite{arora2009computational}) asserts that every ``reasonable'' model of computation can simulate one another with at most polynomial overhead. Consider a multi-tape Turing machine which halts in $T(n)$ steps on inputs of size $n$. This may be simulated on a two-tape Turing machine in $O(T(n) \log T(n))$ steps, or on a one-tape Turing machine in $O(T(n)^2)$ steps \cite{etde_5635755}. One of the most general but still ``reasonable'' models of computation is the random-access machine (RAM), defined in the image of real computers built in the von Neumann architecture. It has an arbitrary number of registers and a simple instruction set. If a RAM takes $T(n)$ steps, this can be simulated on a multi-tape Turing machine in $O(T(n)^3)$ steps \cite{DBLP:journals/jcss/CookR73}.

The explication to $\P$ is then a very convenient choice, in the sense it appears machine independent. The class $\TIME[n^2]$ may be different for different machines, but every ``reasonable'' machine has the same $\P$. Conditioned on this assumption, the study of feasibility can proceed without complex programming arguments as long as one is not too specific of the time complexity of a constructed algorithm. It is a useful assumption. Using the extended Church-Turing thesis, the working complexity theorist is allowed to style their proofs to much the same standard that computability theorists did, using the Church-Turing thesis. 

What constitutes a ``reasonable'' machine is often left vague and unspecified in the literature. Some choice phrasings include ``physically realizable'' and ``realistic''. Any complexity measure that is not proportional to a realistic time of a physical machine is most likely considered unreasonable. Nondeterminism is not immediately assumed to be reasonable. Some models of computation allow for example, the addition of arbitrarily large numbers in constant time, yet real computers appear to take longer to add larger numbers. Such models are also considered unreasonable. We will elaborate on the difficulty of making this ``reasonable'' notion more precise later on in section \ref{ectt2}.

\section{Feasibility as Predicativity and Constructivity}
\label{sec:feasibility_explications}

In this section, we present a philosophical survey of various ways in which feasibility could be explicated, although not always consciously so. They have appeared across different fields and devices of mathematics. When considering feasibility in computation, the most common and accepted explication is to identify feasible computation with the complexity class of $\P$, and the difficulty to provide a principled, let alone conclusive, justification thereof, is well known. If one moves away from the standard Turing machine model of computation, and begins considering complexity in other models, one has come to the subject of the study of complexity, not through asymptotic limitations on Turing Machine resources but through other qualitative characterizations. For example, if one were to consider the equivalence of $\P$ in terms of recursive functions, one might come across the axioms of predicative recursion, which have an arguably more intuitive and philosophical justification grounded on the notion of predicativity. This led us to consider the relationship between predicativity and feasibility, providing a potential explication of feasible computation. When one moves instead onto thinking about feasible reasoning, we enter the field of bounded arithmetic, the study of logical theories whose powers are calibrated to express certain complexity constraints.

Methodologically, in order to frame the discussion in a more formal way, we encourage the use of the tripartite framework of conceptual explication as suggested by De Benedetto \cite{de2021explication}. For each approach, the explication is decomposed into three distinct stages:
\begin{enumerate}
    \item \textbf{Explicandum ($E$):} The vague, pre-theoretical, informal notion of ``feasibility''.
    \item \textbf{Intermediate Semi-formal Explicatum ($I$):} The conceptual bridge, which is a narrower, more rigorous philosophical or mathematical concept used to translate the explicandum into formalizable terms.
    \item \textbf{Formal Explicatum ($F$):} The precise logical or mathematical system resulting from the explication.
\end{enumerate}

Each subsection can be seen as employing such a three-step procedure implicitly. Formally, we proceed through investigating the historical explicata. For each such explicatum, there is both an explicandum and hidden intermediate philosophical bridge which targets the explicatum. We extract these interpretations of feasibility as essentially the same process as in \cite{de2021explication}.

\subsection{Feasibility as Predicativity}

\subsubsection{Predicativity and Predicativism}
\label{sec: predicativity}
Before drawing the connection between predicativity and feasibility, we first provide an introduction to the concept of predicativity and predicativism. In the most familiar form, a definition is said to be \emph{impredicative} if it defines an entity by means of a totality to which that very entity belongs; otherwise it is \emph{predicative}. Historically, this thought developed in two closely related forms. One was Russell’s diagnosis of impredicativity as a form of vicious circularity: a definition is meaningless or empty when it presupposes, in its range of quantification or in its conditions of application, the very object or totality it is meant to determine. The other is Poincaré’s emphasis on invariance or definitional stability: a collection is predicative only if its membership criterion is not altered by the subsequent introduction of new elements into the collection \citep{crosilla2016constructivity}. For example, the maximum of a collection of numbers varies with respect to addition of new elements as it depends on the totality of the collection, and it is for this reason we cannot have a stable definition of maximum for an unfinished class, say for natural numbers, if we think of it as a set that is potentially infinite. Following this example, as \cite{linnebo2023predicativism} suggests, these two stances can be read as naturally connected under a potentialist perspective. If a domain is understood not as a completed infinity but as something extendible through an open-ended process, then circularity is objectionable precisely because it destroys definitional invariance under that process: the admission of new elements may retroactively affect what earlier definitions were supposed to determine.

Although the original discussion over predicativity, as suggested, is closely related to the debates over infinity, it is possible to obtain from the above formulation a more abstract notion of predicativity that applies to any procedure in general. In previous formulations, what causes the definition to be ineffective is that it depends on something that is constantly extending or changing. Since that extension cannot be finished in the case of potential infinity, the definition also cannot finalize to a stable object. But in the potentialist view, an infinite set is just a unceasing extension. Therefore, any extending collection, with an extension procedure that can be performed indefinitely, can be viewed as a generator of infinity. For example, consider a recursively defined sequence, in which case a predicative definition might just determine the limit of some formula defined over the first $n$ terms, although the converse is not always true. In this way, we can decide if a certain definition is predicative, given any recursive procedure. This broad reading of predicativity will become important in our discussion later on.

Currently, the connection between predicativity and feasibility in the literature remains rather vague and unexplored. Leivant is among the first to appeal explicitly to predicativity in order to delineate feasibility. Starting from a view of infinite totalities as evolving rather than completed, he argues that set-existence should be governed only by a \emph{predicative comprehension} principle. The idea is straightforward: if a totality is not given as complete, then its complement cannot be justified, since it presupposes that non-membership is already settled over a completed domain. Unrestricted quantification is likewise objectionable to him insofar as it suggests exhaustive access to the whole universe of elements. He therefore restricts comprehension to positive (unnegated) quantifier-free formulas, and the resulting second-order formalism \(L_2(QF^+)\) turns out to characterize deterministic polynomial time \citep{leivant1994foundational}. In this way, Leivant shows how a potentialist understanding of predicativity can motivate a formalism whose provably convergent programs are exactly the polynomial-time functions. This sets the stage for Bellantoni and Cook, where the same foundational intuition, the separation of completed and incomplete quantity, is carried over into a more concrete theory of recursive definition. However we want to show that the connection between feasibility and predicativity can be stated more explicitly besides a simple binary classification of completed and incomplete variables.

\subsubsection{Recursion-theoretical context}

If feasibility is to be explicated in predicative terms, the next step is to identify a computational analogue of predicative set formation. A natural place to look is the tradition that characterizes classes by restricting recursive schemata. The baseline here is the class of primitive recursive functions ($\mathsf{PR}$), generated from a fixed stock of initial functions---namely the zero function, the successor function \(S(x)=x+1\), and the projection functions---and closed under composition and primitive recursion.

\noindent\textbf{Definition 1.1} (\textit{Primitive Recursion})
If \(g\) and \(h\) are in the class, then so is \(f\), where
\begin{align*}
    f(0,\bar{x}) &= g(\bar{x}), \\
    f(S(n),\bar{x}) &= h\big(n,\bar{x},f(n,\bar{x})\big),
\end{align*}
\noindent for all \(n \in \mathbb{N}\) and all tuples \(\bar{x}\).
The function \(f\) is said to be defined from \(g\) and \(h\) by primitive recursion.

Primitive recursion with the strength of the unbounded minimization operator $\mu$ gives general recursion, which is proven famously to be equivalent to the whole of Turing-computable functions. It is then natural to consider if primitive recursion itself as a restriction of computability can be treated as the explicatum of feasibility. This is, however, quickly refuted due to the fact that primitive recursion, or even its strict subset, the elementary functions, contains functions computable in time $O(\underbrace{2^{2^{\cdot^{\cdot^{n}}}}}_{i \text{ times}})$ for any $i$. 
Since the initial functions are more or less obviously feasible, the inclusion of infeasibly computable functions in $\mathsf{PR}$ must come from the specific species of composition and recursion used. In search of a subset to explicate feasibilism towards, a natural place to look is for more restricted versions of composition and recursion. In terms of a computational procedure, primitive recursion already guarantees well-founded descent on the recursion argument, but it does not yet impose any restriction on how recursively computed values may be reused in later constructions.

The classical formulation of feasible functions, or at least so under the Cobham-Edmonds thesis, is Cobham’s machine independent characterization of $\mathsf{FP}$, which can be read as showing more precisely where primitive recursion exceeds feasibility \cite{Cobham1965-COBTIC}. To weaken from $\mathsf{PR}$ to $\mathsf{FP}$, Cobham alters both the initial function basis and the recursion scheme itself.

Because modern forms of Cobham’s algebra operate over binary representations rather than unary numerals, the standard successor function is replaced by two binary successors, \(s_0(x)=x0\) and \(s_1(x)=x1\), which append a \(0\) or \(1\) to a bit string. More importantly, Cobham introduces a specific fast-growing initial function, the \emph{smash function}, defined by \(x \# y = 2^{|x| \cdot |y|}\). Its role is to provide sufficiently large polynomial bounds to justify the application of his restricted recursion rule. The primitive recursion scheme is then replaced by \emph{Bounded Recursion on Notation}.

\noindent\textbf{Definition 1.2} (\textit{Bounded Recursion on Notation})
For \(i\in\{0,1\}\), if \(h_i\), \(g\), and \(j\) are in the class, then so is \(f\), where
\begin{align*}
    f(0,\bar{x}) &= g(\bar{x}), \qquad\\
    f(s_i(y),\bar{x}) &= h_i\!\big(y,\bar{x},f(y,\bar{x})\big)\ \text{ for } s_i(y) \neq 0,
\end{align*}
\noindent provided that \(f(y,\bar{x}) \le j(y,\bar{x})\) for all \(y,\bar{x}\).\\

\noindent Specifically, this imposes two restrictions on \textbf{Def.~1.1}:
\begin{enumerate}[label=R\arabic*]
    \item \textbf{(Depth Bound)} The ``on notation'' part: recursion depth is bounded by the input length \(|y|\), rather than by the numerical value \(y\).
    \item \textbf{(Size Bound)} The ``bounded'' part: the output length of the resulting function is bounded by some monotone polynomial \(b_f\) of the input length, so that \(|f(\bar{x})| \le b_f(|\bar{x}|)\).
\end{enumerate}

Each restriction is indispensable. If one keeps R1 but drops R2, then composition can reintroduce exponentially large counters into the recursion argument, thereby undoing the length-based restriction. If one keeps R2 but drops R1, the output length remains polynomially bounded, but the recursive unfolding itself may still be primitive-recursive in a much stronger sense. Thus feasibility requires both a restriction on the recursion schema and a restriction on growth under composition.

\subsubsection{Predicativity in Recursion Theory}
Following the predicativist view from Leivant's program, Bellantoni and Cook, by contrast, seek a way of controlling growth without introducing explicit bounds such as R2 or relying on the smash function. As they put it, the question is whether Leivant’s program of using predicativity to characterize computational complexity can be carried over directly into a functional setting, without appeal to logic or provability \cite{10.1145/129712.129740}. Their answer is to internalize the relevant restriction syntactically by distinguishing between \emph{normal} and \emph{safe} arguments.

\noindent\textbf{Definition 1.3} (\textit{Predicative Recursion on Notation})
For \(i\in\{0,1\}\), if \(h_i\), \(g\), and \(j\) are in the class, then so is \(f\), where
\begin{align*}
  f(0,\bar{x}\,~;~\,\bar{a}) &= g(\bar{x}\,~;~\,\bar{a}),\\
  f(s_i(y),\bar{x}\,~;~\,\bar{a}) &= h_i\!\big(y,\bar{x}\,~;~\,\bar{a},\,f(y,\bar{x}\,;\,\bar{a})\big)\quad\text{for } s_i(y)\neq 0,
\end{align*}
Note that the recursive value \(f(y,\bar{x}\,;\,\bar{a})\) only appears in a \emph{safe} position.

\vspace{0.75ex}
\noindent\textbf{Definition 1.4} (\textit{Safe Composition})
If \(h,\bar r,\bar t\) are in the class, then so is $f$ where
\[
  f(\bar{x}\,~;~\,\bar{a}) \;=\; h\big(\,\bar r(\bar{x}\,;)\,~;~\,\bar t(\bar{x}\,;\,\bar{a})\,\big),
\]
where \(h,\bar r,\bar t\in B\); here \(\bar r(\bar{x}\,;)\) is a tuple of functions depending only on the \emph{normal} inputs, while \(\bar t(\bar{x}\,;\,\bar{a})\) may depend on both normal and safe inputs, but its outputs are used only in \emph{safe} positions. These clauses formalize a single underlying idea:
\begin{enumerate}[label=R\arabic*, start=3]
    \item \textbf{(Predicativity)} Recursively computed values may be used as data, but not as fresh control for further recursion. By \textbf{Def.~1.3}, the output of a previous recursive step may occur only in safe position; by \textbf{Def.~1.4}, safe data may not flow back into normal position. Hence the depth of recursion depends only on normal arguments, and recursive output cannot be recycled into new recursion depth.
\end{enumerate}

Both Cobham’s and Bellantoni--Cook’s formulations preserve the ``on notation'' restriction, that is, recursion on input length rather than on input value. Otherwise, a procedure linear in the value of the input would already be exponential in the length of the input, which would immediately violate \(\FP\). The difference lies in how the second restriction is enforced. Cobham does so by explicit bounding; Bellantoni and Cook do so by a predicative discipline on the flow of information inside recursive definitions.

The relevant notion of predicativity can now be restated in explicitly processual terms. Given the recursively computed sequence from our predicative recursion, as suggested in \ref{sec: predicativity}, we are able to decide if a definition is predicative relative to the recursion. 
To bring in a complexity concern, we may define a complexity measure $l$, to measure the length of the recursion. (Formally, it can be defined to read off the safe variable, etc.). This will provide an explication of feasibility: a recursive function is feasible if and only if $l$ on that function is predicative in the classical sense. 

This explication provides a way to formalize the idea that a self-amplifying control procedure is infeasible.

\subsection{Feasibility as Bounded Constructivity}

This section turns from the predicative control of recursion to a different family of explications of feasibility. The common thought here is broadly constructivist: an existential mathematical claim is legitimate only insofar as it is backed by a construction or witness. What distinguishes this family from constructivism in general is that the relevant notion of construction is further restricted by feasibility. The question is no longer merely whether an object can be obtained in principle, but whether its existence can be established by bounded, surveyable, and feasibly verifiable means. In this sense, the guiding idea of the present section is \emph{bounded constructivity}: mathematical reasoning is feasible only when its existential commitments are supported by constructions whose use in proof remains within definite feasible limits.

Within this general perspective, Parikh, Buss, Cook, and Li may be read as articulating different moments of the same idea. Parikh first raises the problem of \emph{feasible existence} inside arithmetic: if a proof asserts the existence of an object whose construction outruns any feasible procedure, in what sense is that existence mathematically significant? Buss develops this pressure into a more systematic proof theory of bounded reasoning, in which the logical form of admissible proofs is calibrated so that existential claims admit feasible witnesses. Li, finally, does not introduce a new technical system, but offers a distinct philosophical reconstruction of Cook's $\PV$, by asking what informal conception of feasible mathematics such a formalism is meant to capture. Accordingly, the present section treats bounded arithmetic primarily through Parikh and Buss, and introduces $\PV$ only insofar as it serves as the formal target of Li's thesis.

\subsubsection{Computation and Reasoning}
\label{computationandreasoning}
Before turning to feasible reasoning in the stricter sense, it is useful to distinguish three levels at which computation and reasoning are related. The connection is not exhausted by the claim that proofs can be checked by a machine. Rather, it becomes progressively tighter as one moves from formal derivation to constructive proof, and finally to bounded construction.

At the first level, formal reasoning is mechanically analyzable because a proof is a finite syntactic object and a derivation is a sequence of rule-governed symbolic steps. In this sense, reasoning is computationally accessible from the outside. Once a deductive system is fixed, proofs can be checked, encoded, enumerated, and in principle searched by mechanical means. This is the level most directly relevant to the Church--Turing thesis and to Kripke's proof-theoretic reconstruction of it: the point is not yet that proofs are themselves computations in any strong sense, but that mechanical calculation can be externalized as formal derivation.

At the second level, the connection becomes internal. In the intuitionistic setting, a proof is not merely a certificate of derivability, but a construction that realizes the proposition proved. Proofs of conjunction provide both components, proofs of disjunction select and prove a disjunct, proofs of existence provide a witness, and proofs of implication behave functionally. This is the setting in which the Curry--Howard correspondence becomes possible. The relevant distinction is therefore not that intuitionistic proofs are machine-checkable while classical proofs are not; proofs in either setting are finite formal objects. The difference is that intuitionistic proofs carry computational content by their very proof-theoretic form.

At the third level, one arrives at feasible reasoning. Once proof is understood constructively, one may ask not only whether a witness or procedure exists, but whether it can be obtained within feasible bounds. Feasible reasoning is thus best understood as bounded constructivity: proofs should not merely establish existence, but should do so in a way that keeps witnesshood, search, or verification within definite resource limits. This is the point at which bounded arithmetics, such as $\PV$ and related theories enter. They should not be viewed simply as intuitionistic systems, but rather as formal attempts to impose feasibility constraints on the constructive content of proof. As we will see later, Parikh's bounded reasoning is motivated by ruling out what he considered infeasible reasoning.

Accordingly, the relation between computation and reasoning in the present discussion may be organized as follows: formal reasoning is mechanically executable as syntax; intuitionistic reasoning is computationally meaningful as construction; and feasible reasoning is a further restriction in which such construction is required to remain within feasible bounds.

\subsubsection{Reasoning in Bounded Arithmetic}

Parikh's 1971 paper marks an important shift in the discussion of feasibility. On the computational side, the question had been to characterize which functions should count as feasible. Parikh relocates the issue to the logical side: which forms of existence and induction should count as feasible within arithmetic? In this respect, his paper may be read as a feasibility-theoretic radicalization of the constructivist concern with mathematical existence. Constructivism requires that existence claims be backed by proof or construction; feasibilism presses the further question of whether such proof or construction remains within feasible bounds.


Parikh's point of departure is exponentiation. Although exponentiation is primitive recursive, he argues that it should not be treated as feasible in the same sense as addition and multiplication. The issue is not merely that exponentiation grows too quickly in practice. More deeply, it gives rise to existence claims whose witnesses are no longer realistically surveyable or obtainable. This is already enough to show that primitive recursiveness is too permissive if feasibility is the target notion \citep{parikh1971existence}.

Parikh also gives a structural reason for singling out exponentiation. Its usual recursive equations determine it uniquely over the standard natural numbers, but not over nonstandard models. As Buss later emphasizes, this separates exponentiation from addition and multiplication in a model-theoretically significant way. Exponentiation is therefore not simply a large function; it is also less tightly controlled by the underlying arithmetic structure. In this sense, it serves as a boundary case revealing that ordinary arithmetic validates forms of existence that exceed any plausible notion of feasible mathematics \citep{buss1999bounded}.


Parikh does not stop with exponentiation as an isolated example. He asks where this excess comes from. At this point he turns to Poincaré's criticism of the justification of induction \cite{poincare1918science}. The point of that criticism is not that induction should be abandoned, but that it cannot simply be reduced to ordinary logical law. If one therefore reflects on the familiar informal justification of induction, one obtains the following picture: one has \(A(0)\); one repeatedly applies \(A(n)\to A(n+1)\), obtaining \(A(1), A(2), \dots\); and one then passes to \((\forall n)A(n)\). The crucial difficulty lies in the unanalysed ``and so on,'' that is, in the passage from a finite progression to the universal conclusion \citep{parikh1971existence}.

Parikh's diagnosis is that, once this passage is made explicit, it is naturally reconstructed in proof-theoretic terms. What is iterated is not merely the truth of successive instances, but the formation of proofs of those instances. In this way, the induction step is read as a transformation from a proof of \(A(n)\) to a proof of \(A(n+1)\). The relevant predicate then becomes
\[
B(n): \text{ there exists a proof of } A(n).
\]
Accordingly, Parikh takes the ordinary justification of induction to reduce arbitrary induction to induction on existential formulas. Whether this is read as a strict formal reduction or as a diagnosis of the justificatory structure is less important than the conclusion he draws from it. The real source of unbounded strength lies in induction on existential predicates. He immediately adds that primitive recursion within Peano arithmetic is justified in the same way. The philosophical consequence is that the amplification of existential commitments occurs at exactly this point, just as unrestricted recursion is the corresponding point of amplification on the computational side. Exponentiation is only the clearest symptom. The deeper source of infeasibility is unrestricted induction---more precisely unrestricted induction as it operates through existential formulas.


At this stage Parikh's argument is still mainly negative. It identifies where the difficulty lies, but it does not yet by itself derive the precise form of the repair. His positive proposal is the class of bounded formulas, later written \(\Delta_0\), and the corresponding theory \(PB\), now standardly denoted \(I\Delta_0\). Syntactically, these are formulas in which every quantifier is explicitly bounded by a term, and \(I\Delta_0\) is obtained by restricting induction to that class \citep{parikh1971existence}.

Historically, Parikh does not formulate a single explicit philosophical principle from which boundedness is deduced. Rather, the paper combines the negative diagnosis above with a positive appeal to a more ``concrete'' fragment of arithmetic and with a corresponding computational picture. For this reason, the move from unrestricted induction to bounded induction should not be presented as a proper deduction from the earlier argument. It is better understood as Parikh's proposed repair once arbitrary existential induction has been rejected.

The present interpretation is that what unifies these moves is a principle of bounded existential legitimacy. If feasible mathematics must control existential commitments, then induction may be retained only where the relevant witness range is already internally delimited. On this reading, bounded formulas are not introduced merely because they suggest finite search, but because they restrict existential reasoning to cases in which the search for a witness remains tied to antecedently given arithmetic bounds. This also explains why \(I\Delta_0\) should be viewed primarily as a theory of bounded induction rather than merely as a theory of bounded quantification. Buss's retrospective discussion is helpful here not because it supplies a new foundation, but because it shows more clearly what the bounded fragment excludes and why exponentiation remains the paradigmatic case \citep{buss1999bounded}.

Parikh's theorem confirms that this restriction is not merely syntactic.
If $A$ is bounded and \(I\Delta_0\) proves \((\forall x)(\exists y)\,A(x,y)\), then the theory already proves such a statement with a polynomial bound on \(y\). Thus, provable existence in bounded arithmetic carries bounded witnesshood. In this sense, Parikh's contribution is to relocate the problem of feasibility from the growth of functions alone to the logical form by which arithmetic amplifies existential commitments, and to propose bounded induction as the first systematic repair.

\subsubsection{Later Development and Connection to Complexity Theory}
If Parikh's contribution is to identify the proof-theoretic source of infeasible existence and to introduce bounded induction as a first repair, Buss's contribution is mainly systematic rather than foundational. He does not supply a wholly new philosophical principle for boundedness. Instead, he develops bounded arithmetic into a more finely articulated framework and connects its fragments with computational complexity \citep{buss1985bounded}.

The crucial point is that \(I\Delta_0\) does not remain the final or uniquely canonical formalization of feasible reasoning. In Buss's work, bounded arithmetic is reorganized into stronger and weaker theories, especially \(S_2\) and its subtheories \(S^i_2\) and \(T^i_2\). In this setting, one no longer asks only whether induction has been restricted to a bounded fragment. One also asks how different bounded proof principles correspond to different computational strengths.

From the present point of view, \(S^1_2\) is significant because it belongs to this finer calibration of feasible reasoning. It should not be identified with Parikh's original \(I\Delta_0\), but understood as part of a later bounded-arithmetic hierarchy designed to track more specific complexity-theoretic notions, especially those surrounding polynomial-time reasoning. The conceptual shift is therefore from repair to stratification: Parikh isolates a bounded fragment in response to the excesses of unrestricted induction, whereas Buss turns bounded arithmetic into a hierarchy of theories whose strengths can be compared with complexity classes, in particular, the levels of the polynomial-time hierarchy $\mathsf{PH}$.

For present purposes, the importance of Buss lies in this systematic extension. He shows how the bounded-arithmetic idea can serve not only as a negative response to infeasible existence, but also as a framework within which different levels of feasible reasoning may be logically distinguished.

\subsubsection{Li: Feasible Mathematics and \texorpdfstring{$\PV$}{PV}}
Li's recent discussion has a different role from Parikh's and Buss's. His concern is to step back from systems such as Cook's \(\PV\) and ask what informal picture of feasible mathematics they are meant to formalize \citep{li2025introduction}.

The starting point of Li's proposal is explicitly constructivist. He begins from a BHK-style reading of proof, under which an existential statement requires a uniform procedure that constructs a witness together with a proof of the relevant property, and a universal statement requires a uniform procedure that generates proofs of its instances. The crucial further move is to reinterpret the relevant notion of effective procedure as polynomial-time computation. In this way, feasible mathematics is introduced as a complexity-bounded version of constructive mathematics. The issue is no longer merely whether a mathematical object can be constructed, but whether it can be constructed feasibly.

This point is important for the status of Li's three postulates. He does not derive polynomial time from certain philosophical principles. Rather, polynomial time is fixed in advance as the intended notion of feasibility, and the postulates are meant to describe what a corresponding mathematical practice should look like. In this sense, Li's Feasible Mathematics Thesis may be read only as a proof-theoretic analogue of the Cobham-Edmonds thesis. Just as the latter identifies the informal notion of feasible computation with polynomial-time computation, the former claims that the corresponding informal notion of feasible mathematical proof is captured by \(\PV\).

The first postulate concerns feasible functions. Its point is not merely that feasible functions should be admitted, but that they should be admitted only when their feasibility is \emph{clearly demonstrated} by their construction. This gives the postulate a stronger epistemic content than a bare appeal to the class $\FP$. A function is not accepted simply because it happens extensionally to be polynomial-time computable; rather, it must come equipped with a construction that makes its feasibility mathematically transparent. The foundational idea here is therefore one of \emph{displayed} or \emph{warranted} feasibility, not mere extensional classification.

The second postulate concerns definition axioms. Once a function is admitted because its construction clearly demonstrates feasibility, the mathematical facts encoding that construction should themselves be available as axioms. Otherwise one could compute with the function but not reason about it. This postulate therefore internalizes feasible construction into mathematics proper: the construction is no longer merely an external algorithmic fact, but becomes part of the proof-theoretic resources of the theory. Li explicitly describes these axioms as capturing the local behavior of the function, even if they are not by themselves sufficient for more global arguments.

The third postulate concerns structural induction. Here Li arrives at an issue close to Parikh's, but from the positive side. Local definition axioms do not suffice to prove global properties, so some form of induction is required. However, not every induction principle is feasible. Li's restriction is that induction is permitted only for predicates whose truth remains feasibly verifiable; in his austere formulation, this becomes induction over the binary representation of the input for equations of the form \(g(x)=h(x)\), where \(g\) and \(h\) are already admissible functions. The underlying thought is that the chain of reasoning must itself remain polynomially bounded. Thus, the postulate does not merely add induction to the system; it articulates a specifically feasible notion of global reasoning.

Taken together, the three postulates express a unified conception. Feasible mathematics is that in which admissible constructions must display their feasibility, those constructions must be internalized as definition axioms, and global reasoning over them must proceed only by induction principles whose verification remains itself feasible. In this respect, Li's proposal is best understood as an explication of a style of mathematical practice---that of the humble programmer, as defined by Dijkstra \cite{Dijkstra1972}---rather than as a purely formal criterion. The central issue becomes which constructions, axioms, and proofs can be responsibly used by a feasible mathematician.

This also explains the role of \(\PV\). Cook's theory \(\PV\), taken by itself, largely presupposes rather than explicates the underlying notion of feasibility, since the target class of functions is fixed in advance \cite{Cook1975}. Li's thesis is that \(\PV\) should nevertheless be seen as the correct formal realization of the broader informal picture given by the three postulates.\citep{li2025introduction}.

\section{Feasibility as Physical Limitations}

We now present a distinct source of feasibility from that which is derived as a form of mathematics, but we first discuss the effective case. An interpretation of computation is that which is implemented by computing machines which are potentially realizable by fully mechanical methods. Once a device has been actualized, it is obviously realizable. By realizable, we mean only that one could hypothetically construct it without necessarily expending the effort to do so. When a device of any sort is first imagined, it is considered realizable if its existence is consistent with our sense experience of the world we inhabit. Being vague, imprecise, subjective, and deceivable, our sense experience is not at all a solid foundation to build upon. For example, the Aristotelian fallacy that heavier objects fall faster was well accepted for nearly two millennia until Galileo's experiments. Rather than sense experience, we should instead delegate to the laws of physics. What is a physical theory if not just an objective universal generalization of a necessarily finite collection of sense experiences? This relationship between physical theories and sense experience is one of the oldest and most well studied problems in the philosophy of science, and our framework will inherit at least two limitations rooted in that existing work. First, no scientific theory, let alone a physical theory, can ever be proven. Increasing amounts of evidence can never confirm a theory. Second, there may be distinct physical theories which are all wholly satisfactory. What criterion of adequacy (for example, Popper's falsifiability \cite{popper2005logic}) we should take in choosing a theory is also a difficult question. Despite these points, we shall mark for our use that the \textit{laws of physics} are the consistent union of all well accepted physical theories which can be used to explain and predict matter, energy, time, and so forth. We then shall define a device to be realizable if its existence is consistent with these laws of physics.

\subsection{Grounding Explication in Physics}

An elementary example of an application of our definition is that the non-realizability of a perpetual motion machine follows from the conservation of energy. Inconsistency with the laws of physics can be understood as a kind of undecidability result, conditioned on those laws of physics being correct. If a device ``computes'' a non-computable function, then it should not be realizable. But this is an undecidability notion, not an infeasibility notion. We wish to assign meaning to what it means for there to be a ``feasible device'' which may only compute a feasibly computable function. We should first consider what resources there are to measure of a physical machine. Some candidates include space-time volume, energy, mass, initial stored energy, and the maximum magnitudes of all physical variables. Rather than commit to any combination of these, a feasible machine should be feasible \textit{in all resources}, to be a true explicandum of feasibility. Tradeoffs between resources are common in engineering---decreasing one at the cost of more than increasing another. In the intent of feasibility, this should be disallowed.

The perspective that feasibilism should be interpreted as feasible in all resources also resolves a fundamental question of the Cobham-Edmonds thesis. The class $\P$ is a time complexity class. Why do we most associate feasibility with time complexity, as opposed to any other complexity measure? Time is not necessarily intrinsically special, it simply majorizes all other non-static complexity measures. Consider for example, the relationship between space and time complexity of a Turing machine. At a certain computation step, the machine can choose to access a previously used cell (increasing the time, but not the measured space), or the machine could access a new cell (increasing both the time and space). In both cases, the time will increase, but the space may not. This argument generalizes to any non-static complexity measure, including measures not immediately obvious to an observer. Static complexity measures are a separate interest, such as in Kolmogorov complexity. However, in asymptotic analysis we study the behavior of algorithms when executed on inputs whose lengths far exceed the length of the program. In this case, the time will also majorize static measures. 

There are a few physical phenomena which are good candidates for feasible and infeasible physical processes. Consider any ``irreversible'' thermodynamic process. Baking a cake may require some fixed amount of energy, but unbaking a cake seems much more difficult. Reversing the process of baking a cake to its singular initial microstate (out of exponentially many possible initial microstates) would require a seemingly infeasible amount of energy. This is an infeasibility notion which is distinct from the impossibility notion which disallows the perpetual motion machine.

\subsection{Discrete Physical Systems and Computability}

Turing was primarily concerned only with idealized mental procedures. In his thesis, he imagines the work done by a human computor, and explicates this action to that of a Turing machine. The increased prevalence of high speed machine computers only occurred many years after the publication of Turing's 1936 paper \cite{Turing1936-TUROCN}. Much later, Robin Gandy sought a formal analysis of machine computation, like his advisor, Turing, had studied human computation \cite{Gandy1980}. 

Gandy first gives an explicit definition of what he calls a ``\textit{discrete deterministic mechanical device}'', and what has now been referred to as a Gandy machine \cite{shagrir2022nature}. A Gandy machine is a discrete-time system with states drawn from a finite-dimensional configuration space, satisfying four principles:

\begin{enumerate}
  \item Devices and parts of the machine can be labeled by hereditary finite sets.

  \item Labels of parts of the device must be finitely bounded. You cannot recurse on smaller subassemblies infinitely.

  \item Given labels of a device, there exists a unique manner to reconstruct the device.

  \item (\textit{Principle of Local Causality}) Let $x$ be a configuration of the device at some intermediate step, and $x'$ the next immediate configuration of the device. For $s'$ a region of $x'$, there exists a finite neighborhood, $s$ of $x$ such that $s'$ only depends on $s$. A region may only be influenced by changes in its finite neighborhood.
\end{enumerate}

Each principle is explicitly grounded from a physically motivated limitation. First, there is a lower bound on the density of information: only finitely many distinguishable structures may occupy a bounded region of space. This forces the content of the first three conditions---a machine must be assembled from finitely many identifiable parts of bounded size, arranged in a hierarchy of bounded depth, and with no infinitely descending chains of sub-assemblies. Second is an upper bound on the velocity of propagation of information, which Gandy draws from the geometry of space-time: contemporary physics disallows instantaneous action at a distance. From these principles, every Gandy machine's transition function is computable by a Turing machine, so any discrete physical system satisfying the constraints may be simulated on a Turing machine. Gandy's principles are extremely general, and it is difficult to imagine a realizable device which does not satisfy these principles. 
While Turing made the appeal to the sense experience of a human computor to assert that only a bounded portion of the state may be changed in a single step, Gandy---concerned instead with discrete mechanical devices rather than human computation---justifies this same property with an appeal to the theory of relativity which he titles ``\textit{the principle of local causality}''. Without this fourth principle, the simulation of a Gandy machine on a Turing machine cannot proceed. It is necessary for his explication.

The framework provided by Gandy, which explicates discrete machine computation restricted by physics to the Turing machine, will later serve as an explication of feasibilism, as we shall explain in section \ref{gandyfeasible}. 

\subsection{Complexity of Analog Physical Computers}
Gandy's work only explicitly discusses discrete devices, so we discuss the analog case separately and briefly. It is worth mentioning since there exists some criticism of this portion of Gandy's analysis \cite{10.1023/A:1008236522699,copeland2007physical, shagrir2022nature}, although their complaints are not applicable to our later discussion of Gandy's thesis in the context of feasibilism. Analog computers have been actualized for some time; the simplest example is that of a ball-and-disc integrator. A precise disc is spun at a fixed rate, and a large precise bearing is moved across the surface of the spinning disc according to some function $f$. The spin of the bearing is physically translated to an output shaft, whose movement encodes the value of the definite integral of $f$. We mention this example because it is not obvious at all if the physics which govern this kind of computation permit an obvious discrete simulation. 

The complexity of analog computation has been best studied by Vergis, Steiglitz, and Dickinson \cite{Vergis1986}. They formalize a nonuniform deterministic definition of an analog computer with several requirements. Most importantly, they require of their definition that such analog machines have an absolute precision $\epsilon > 0$. This idealization is necessary to make the definition robust. Every idealized physical model of a real system cannot be more sensitive to small changes than the real system is to minor physical perturbations, such as the gravitational pull of the moon, or temperature fluctuations. This use of precision as a resource also ensures devices which satisfy their definition maintain their stability under physical scaling. They formulate an analog version of the extended Church-Turing thesis, which states that any analog machine (as they define) can be simulated by a Turing machine with at most polynomial-time overhead as a function in all resources used by the analog machine. Their main theorem states if there is an analog machine to solve a problem which is conjectured to require exponential time on discrete devices, the analog machine must use an exponential amount of resources. If feasible is to mean feasible in all resources, we may interpret their result to mean we are no less general restricting our study to discrete devices only. The problem they have chosen is not proven to be exponential time to discrete devices, but is only so under well accepted, plausible complexity assumptions. We will discuss arguments of this form in section \ref{complexity}.

\section{Proving the Cobham-Edmonds thesis}
\label{provingthethesis}
There are three particular types of argument for the Church-Turing thesis. The first type is the presentation of the Church-Turing thesis as a falsifiable scientific hypothesis, and the display of evidence in its favor: various definitions of computation which are capable of simulating each other. 

The second and third types of argument are vastly superior. They are rigorous justifications of the explication itself. These are not mathematical proofs, in the sense they can ever exist in a formal system, but they are as well written and as convincing as any other mathematical proof. The second type of argument of this kind is Turing's thesis. The third argument of this kind, more recently, is Kripke's \cite{kripke2013}. Although Kripke's exposition is undeniably the most explicit, it exists in two prior forms. First is Church's attempt to equate the $\lambda$-definable functions with those which are ``effectively calculable'' \cite{church1936unsolvable}. His argument famously appeared deficient to G\"odel \cite{kleene2007origins}. A prior argument of this form is also Turing's argument II. His argument I is now denoted as Turing's thesis.

In the feasible case, the Cobham-Edmonds thesis is only presented in the literature with arguments of the first kind, if it is even made explicit. Although this type of argument is more convincing in the effective case, in the feasible case, the situation is entirely different. The extended Church-Turing thesis states that machines which satisfy some ambiguous criterion of ``reasonable'' are able to feasibly simulate each other. It does not obviously follow from this that $\P$ should be the explicatum, because it is not obvious what is or isn't a ``reasonable'' machine. The effective case does not have this issue because the models of computation are as generalized as possible.

What is needed for the Cobham-Edmonds thesis are arguments of the superior sort. Our contribution here is the observation that under analogous restrictions, arguments of the second and third kind of the Church-Turing thesis can be made more precise to suffice as rigorous arguments of the Cobham-Edmonds thesis. We now present feasibly restricted versions of Turing's thesis, Gandy's thesis, and Kripke's thesis. While their arguments are rigorous and convincing, each of our three feasibly restricted arguments will fall victim to the same flaw. We will later argue this flaw must exist in principle.

\subsection{Turing's Thesis and Feasibilism}
\label{feasibleturingthesis}

We briefly summarize Turing's legendary argument, the Direct Appeal to Intuition \cite{Turing1936-TUROCN}. An alternative presentation is given by Kleene \cite{Kleene1952-KLEITM}. Turing imagines the work done by any human computor, working with pen and paper following some fixed but arbitrary mental procedure. Through a sequence of distillations, he makes the model increasingly more formal, until we simply have the Turing machine.

What is first observed is that the act of computation is independent of the geometry of the writing surface, so the pen and paper is replaced with an unbounded tape. The next simplification is the observation that the number of symbols which may be printed is finite. Having finitely many symbols is not less useful, as one may use sequences of symbols, and treat them as single symbols. Having infinitely many symbols is also not more useful, as most would be indistinguishable from each other. The next observation is that the human computor may only interact with the tape in a local way. There are some bounds on the number of symbols which may be read from and written to in a single step. There is also a bound to the number of cells away the human computor can observe next. More symbols may be read from and written to, but this necessitates successive steps. The pen and eyes of the computor are replaced by a simple tape head, which may scan the symbol, write a new one, and move left or right a finite distance away from its current position. The scanned symbols and current state of mind uniquely determine the next state of mind and the next cells to be observed. There are also only finitely many states of mind, with the same argument as to why there is a finite alphabet. A use of an infinity of states of mind can be avoided by writing more symbols to the tape. The mind of the human computor is then replaced with a Turing machine's transition function and the explication is complete. 

Turing justifies each step by his sense experience of a human computor, invoking an implicit assumption of fathomability, one which only distinguishes finite from infinite procedures. Observe that his explication still holds in the case that restrictions are made to the human computor. During the process, one still receives a formal model, but with explicated restrictions. For example, instead of considering a human computor operating with an arbitrary (but fixed) mental procedure, suppose the human computor was operating the fixed mental procedure of bitwise addition. The output would exactly be a fixed specific Turing machine which computes addition exactly in the method of the mental procedure of the computor. Other models of computation may simulatable on the Turing machine for this reason. We say a model of computation may pass the \textit{fathomability criterion} if a human computor working with pen and paper may simulate the model of computation using a description of it. The $\lambda$-calculus, the G\"odel-Kleene-Herbrand recursive functions, general programming languages, and semi-Thue systems all pass this fathomability criterion. One may employ the argument in Turing's thesis, imagining a human computor performing a fixed mental procedure of the simulation of a model of computation, to explicate out a fixed Turing machine formally simulating that model of computation. The Turing machine itself passes this criterion, which is why there exists a universal Turing machine. It is an important detail, often unstated, that a model of computation must pass the fathomability criterion to be equivalent to the Turing machine. One may imagine all kinds of creative super-Turing models of computation, machines with oracle access to an undecidable language, infinite length programs, Zeno machines, and so on. But in contraposition, if they really are super-Turing, they explicitly must fail this fathomability criterion. Otherwise Turing's thesis would allow them to be simulated on a Turing-machine. 

Turing wishes to consider the most general definition of computation possible so that the explicatum is as general as possible. Again, his explication is successful if one considers a less general model. Suppose we specialize his argument, and replace the requirement of fathomability with one of feasibility. All that is feasible must necessarily also be fathomable, so we should expect restrictions rather than generalizations in the explicatum. What restrictions will be observed? 

Consider a human computor explicitly performing a feasible computation, that is, feasible in all resources. We feasibilize every step of Turing's explication. Originally, the human computor had access to an arbitrary amount of pen and paper, but now they have access only to a feasible amount. We may still replace their paper with a tape of cells, but there can only be a feasible amount of tape. The number of symbols which may be printed is finite. Since all that is finite is already feasible, this step of Turing's thesis is invariant to our restriction. Similarly for the bound to the number of cells which are read and written to in a single step, and the number of states of mind. We may assume the human computor must also only take a feasible number of steps during their computation. Our output is a Turing machine, with a feasible time and space bound. Since using space requires time, we simply have a feasibly time-bounded Turing machine.

Turing's thesis is extremely robust, which is what makes it such an excellent model for computability theory. Small changes, such as allowing the machine to modify finitely many cells at once, move several cells at once, increasing the alphabet by finitely many symbols, are all accommodated during the explication. 

Analogously, the feasible version of Turing's thesis, that which we have presented here, is exactly what also makes the Turing machine a suitable model for computational complexity. The accepted start of the field began when Hartmanis and Stearns gave their definition of the complexity of a problem to be the number of steps a Turing machine variant takes as a function of the input size \cite{hartmanis1965computational}. They note that minor changes to the model of computation do not appear to change the intrinsic complexity of the problems themselves. The robustness of their definition is inherited from the robustness of Turing's thesis. An (apocryphal) story attributed to Hartmanis, is that they first attempted to base their definition on recursive function theory, but found it too complicated, and that the Turing machine made everything so simple \cite{lipton2022hartmanis}.

Our argument is limited. We are able to say that feasible computation must explicate to a time bounded Turing machine, but we are unable to say that this time bound must necessarily be a polynomial in the size of the input. 

\subsection{Gandy's Thesis and Feasibilism}
\label{gandyfeasible}

The framework provided by Gandy---which explicates discrete machine computation subject to physical limitations onto the Turing machine---can be repurposed and modified to serve as an explication of feasibilism. The explicandum under consideration by Gandy was ``\textit{some glorious contraption of gleaming brass and polished mahogany}''. Consider then a feasibly restricted version of this, denoted for our use as a feasible Gandy machine. Such a device, being feasible in all resources, is of a feasible size, and consumes a feasible amount of energy and so on. The justification of Gandy's four principles only follows from two physical assumptions: the lower bound on information density, and the upper bound on signal speed propagation. We shall argue that a feasible Gandy machine is limited by the same four principles, but our justification will instead follow from feasible versions of those two physical assumptions.

We derive a feasible version of the first three principles. Further analysis has shown that Gandy's complex form of description using hereditary finite sets is actually unnecessary \cite{sieg2002mathematical}. We proceed with a more standard argument. Every Gandy machine must have a finite description (otherwise, one could exist to decide the halting problem), so simply consider any banal encoding scheme. All subassemblies and parts are labeled by natural numbers. If the machine is of a feasible size, there can only be a feasible number of these labels. Given these labels, the machine is not only uniquely reconstructible, but it is obviously feasible to do so. The true goal of the form of description is so that the simulation of a feasible Gandy machine may proceed feasibly. It is more difficult to come up with an encoding scheme where the limitations in simulation are due to the encoding strategy of the machine. For example, an encoding scheme of a Turing machine could be a natural number $j$ such that machine $M$ is the $j$th smallest machine when machines are ordered by their shortest descriptions. While this is unique and concise, it is not obvious whether one can feasibly determine any information about the machine, let alone feasibly simulate it. A much better encoding scheme would be the concatenation of tuples of its elements. There is no claim that this is the optimal form of description; however, it is obviously sufficient for both feasible simulation and feasibly determining all static properties of the machine.  

We now require a feasible version of the principle of local causality. Gandy's justification follows from an appeal to the theory of relativity, but this is not sufficient for us. Such extremal limits do not necessarily justify that our feasible Gandy machine is feasibly bounded, only bounded. It is plausible that any feasibly restricted mechanical device need not be limited by the most general models of physics possible. Given resource limitations, it is perhaps the case it can be fully limited by casting it within a simpler, superseded, but feasible model of physics. Although we cannot say too precisely what these laws must be, limitations of speed of the device may prevent relativistic effects from being feasibly measurable, and limitations of precision in manufacturing may remove quantum effects from being feasibly observed. Perhaps these feasible laws of physics are just classical mechanics, or one of the many contentious formulations of discrete physics. A justification without assuming relativistic effects would be more general. It appears that Gandy almost identifies this issue, or one like it. He believes his appeal to relativity is required in order for the simulation of a Gandy machine on a Turing machine to proceed, and thus, Gandy's thesis to be a generalization of Turing's thesis. In idealized Newtonian mechanics, there may be rods of arbitrary length, which may perform instantaneous action at a distance, and he conjectures whether or not there is an idealized Newtonian system to compute a non-computable function. Such a counterexample does not affect our consideration. Our machine being of a feasible size, we would only admit rods of a feasible length. This is sufficient to prevent instantaneous action at a distance, and therefore, we must have local causality.

Armed with his principles, Gandy's next step in his argument is the simulation of a Gandy machine on a Turing machine. This finalizes the Turing machine as the explicatum of Gandy's thesis. We wish to mirror this in the feasible case, but require a more detailed argument to demonstrate the equivalence between a feasible Gandy machine and a feasible Turing machine. First we mention that the Gandy machine is a direct generalization of the Turing machine, and every resource bounded Turing machine may be simulated on a Gandy machine with identical resource bounds. Therefore, every feasible Turing machine is also a feasible Gandy machine. 

Conversely, we wish to show that for every feasible Gandy machine, there is an equivalent feasible Turing machine, so we proceed by a feasible simulation of a feasible Gandy machine on a feasible Turing machine. The feasible limitations will enable a feasible simulation, something that is not true in the general case. For sequential configurations $x,x'$, the feasible Gandy machine may only modify a feasibly bounded amount across the space. Each bounded region is influenced only by changes in its local neighborhood, yes, but the sum over all changed regions must also be feasible. With the configuration itself at most of size $f(n)$, we see the volume of the modified region must be less than this. The Turing machine may take time $O(f(n))$ to compute a single step of the Gandy machine, assuming the Gandy machine's transition function is constant time to compute its application to a constant sized region. If the Gandy machine takes at most $f(n)$ steps, then the Turing machine will simulate this in $O(f(n)^2)$ time and $O(f(n))$ space. Conditioned on $O(f(n)^2)$ being a feasible resource bound whenever $f(n)$ is, this Turing machine is feasible and the explication is complete.

We emphasize that we were only able to conclude the argument with the additional unjustified assumption that if $f(n)$ is feasible then so is $O(f(n)^2)$. Whether or not $O(f(n)^2)$ is necessarily a polynomial, we cannot say since we were unable to justify precise bounds on the Gandy machine.

\subsection{Kripke's Thesis and Feasibilism}
\label{kripke}
We clarify the argument we denote as Kripke's thesis and its relationship to Turing's thesis \cite{kripke2013}. Consider the relationships among the intuitive concepts of deductive thought and computation, and the formal concepts of first-order logic and the Turing machine. 

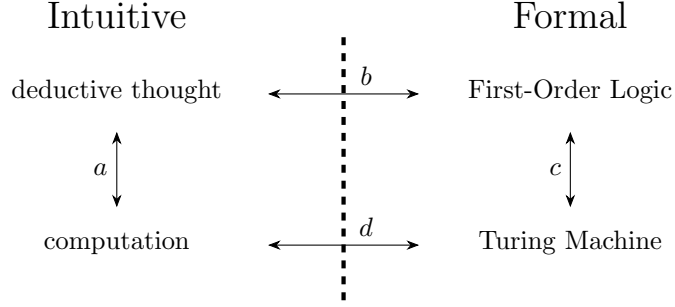
\begin{center}
\begin{tikzpicture}[scale=1]
  \coordinate (A) at (0.00, 5.00);
  \coordinate (B) at (0.00, 1.50);
  \coordinate (D) at (3.00, 5.0);
  \coordinate (C) at (-3.00, 5.0);

  \coordinate (E) at (-3.00, 4.00);

  \coordinate (F) at (-3.00, 2.00);

  \coordinate (G) at (3.00, 4.00);

  \coordinate (H) at (3.00, 2.00);

  \coordinate (I) at (-3.00, 3.75);
  \coordinate (J) at (-3.00, 2.75);
  
  \coordinate (K) at (3.00, 3.75);
  \coordinate (L) at (3.00, 2.75);
  
  \coordinate (dt) at (-1.0, 4.25);
  \coordinate (fol) at (1.0, 4.25);

  \coordinate (com) at (-1.0, 2.25);
  \coordinate (tm) at (1.0, 2.25);
  
  \draw[ultra thick,dashed] (A) -- (B);
  \node[above, font=\Large] at (D) {Formal};
  \node[above, font=\Large] at (C) {Intuitive};
  \node[above] at (E) {deductive thought};
  \node[above] at (F) {computation};
  \node[above] at (G) {First-Order Logic
};
  \node[above] at (H) {Turing Machine
};

  \draw[Stealth-Stealth] (I) -- node[left, text=black, midway] {$a$} (J);
  
  \draw[Stealth-Stealth] (K) -- node[left, text=black, midway] {$c$} (L);

  \draw[Stealth-Stealth] (dt) -- node[above, text=black, pos=0.65] {$b$} (fol);
  
  \draw[Stealth-Stealth] (com) -- node[above, text=black, pos=0.65] {$d$} (tm);
\end{tikzpicture}
\end{center}

We describe these four relationships.
\begin{enumerate}[label=$\alph*$)]
    \item There is a correspondence between computation and deductive thought. Kripke asserts that computation is a special form of deductive reasoning, but we believe that computation is as general as deductive thought, and have represented this as a correspondence. This makes no difference for Kripke's argument. Working with such general notions in the intuitive domain, any such demarcation must be arbitrary and arguable anyway.
    
    \item Hilbert's thesis asserts that the intuitive notion of \textit{provable} may be explicated to the formal notion of \textit{provable in first--order logic}. Since we are corresponding an intuitive notion with a formal one, we can only ask for a thesis, and never a theorem \cite{barwise1977introduction}.
    
    \item A relationship between formal models need not rely on a thesis, and we have a theorem of the correspondence between first-order logic, and the Turing machine. For each computable language $L$ and word $w$, there exists a proof in first-order logic either that $w\in L$ or $w \not \in L$. With $L$ computable, there exists a Turing machine to decide it, and the sequential steps of the proof  in first-order logic encode the sequential steps of the Turing machine. The conclusion of the proof is the assertion that the Turing machine accepts or rejects $w$. Conversely, there exists a Turing machine to output exactly and only all true statements of first-order logic. This machine simply takes the axioms and rules of deduction of first-order logic and computes all syntactically valid proofs and their conclusions. By G\"odel's completeness theorem, those which are true in all models are exactly those which are syntactically provable in first-order logic, thus this machine outputs all true theorems. 

    \item Turing's thesis gives this explication directly, imagining the work done by a human computor, and rigorously arguing there exists a Turing machine for this equivalent action. The converse follows since the Turing machine passes the fathomability criterion. A human computor with pen and paper may simulate a Turing machine. 
\end{enumerate}
While Turing's thesis argues this directly and immediately ($d$), Kripke's thesis is the composition of $a + b + c$. Kripke makes $a,b$ explicit, while Church's argument and Turing's argument II are more focused on $c$. First, Kripke asserts that computation is a special form of deductive thought. Then Hilbert's thesis is conditioned upon, and from there, the formal equivalence between first-order logic and the Turing machine convinces us that all intuitive computation has a formal Turing machine corresponding to it. The unprovable assumption of the Church-Turing thesis may be taken then as a corollary of the unprovable assumption of Hilbert's thesis. 

This presentation allows us to argue more precisely that Kripke's thesis feasibilizes. Consider the relationship between the intuitive notions of feasible mathematics, and feasible computation, and the formal notions of $\PV$ and $\P$. 

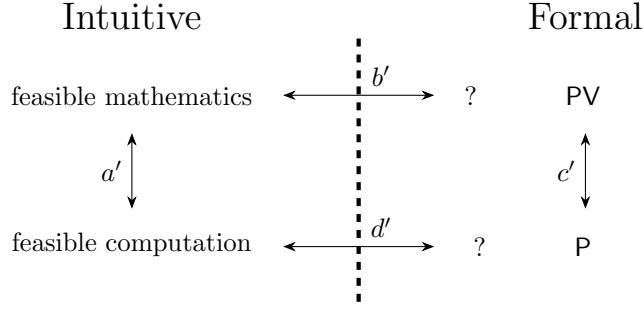
\begin{center}
\begin{tikzpicture}[scale=1]
  \coordinate (A) at (0.00, 5.00);
  \coordinate (B) at (0.00, 1.50);
  \coordinate (D) at (3.00, 5.0);
  \coordinate (C) at (-3.00, 5.0);

  \coordinate (E) at (-3.00, 4.00);

  \coordinate (F) at (-3.00, 2.00);

  \coordinate (G) at (2.30, 4.00);

  \coordinate (H) at (2.30, 2.00);

  \coordinate (I) at (-3.00, 3.75);
  \coordinate (J) at (-3.00, 2.75);
  
  \coordinate (K) at (3.00, 3.75);
  \coordinate (L) at (3.00, 2.75);
  
  \coordinate (dt) at (-1.0, 4.25);
  \coordinate (fol) at (1.0, 4.25);

  \coordinate (com) at (-1.0, 2.25);
  \coordinate (tm) at (1.0, 2.25);
  
  \draw[ultra thick,dashed] (A) -- (B);
  \node[above, font=\Large] at (D) {Formal};
  \node[above, font=\Large] at (C) {Intuitive};
  \node[above] at (E) {feasible mathematics};
  \node[above] at (F) {feasible computation};
  \node[above] at (G) {$?~~~~~~~~~~\PV$
};
  \node[above] at (H) {$?~~~~~~~~~~\P$
};

  \draw[Stealth-Stealth] (I) -- node[left, text=black, midway] {$a'$} (J);
  
  \draw[Stealth-Stealth] (K) -- node[left, text=black, midway] {$c'$} (L);

  \draw[Stealth-Stealth] (dt) -- node[above, text=black, pos=0.65] {$b'$} (fol);
  
  \draw[Stealth-Stealth] (com) -- node[above, text=black, pos=0.65] {$d'$} (tm);
\end{tikzpicture}
\end{center}

Again, we provide justification for our four relationships.
\begin{enumerate}[label=$\alph*'$)]
\item We began this discussion in \ref{computationandreasoning}, but emphasize it here. Conditioned on a definition of feasible computation, our interpretation of feasible mathematics is a feasibly restricted version of the BHK-interpretation of proof. While the BHK-interpretation may simply require a computable procedure, we would further require that this procedure to be feasibly computable. Conversely, given an interpretation of feasible mathematics, we shall admit a model of computation to be feasible only if there exists a true theorem in feasible mathematics whose proof steps encode the computation steps.

\item We seek a feasible version of Hilbert's thesis. In the original definitions of $\PV$, Cook defines the \textit{Verifiability thesis}, but presents no real arguments in its favor \cite{Cook1975}. As an alternative to this, Li proposes the \textit{Thesis of Feasible Mathematics} in section 2.2 of \cite{li2025introduction}, which states that a statement is provable in feasible mathematics if and only if there is a straightforward formalization of the statement in $\PV$ that is provable in $\PV$. Li characterizes his explicandum of feasible mathematics with three postulates, and corresponds this with $\PV$. Both Cook's and Li's theses attempt to connect these notions in an unambiguous way, but they are both unsuitable for our use. Since they both implicitly assume the Cobham-Edmonds thesis, use of their theses to demonstrate the Cobham-Edmonds thesis would be circular. Something may still yet be salvaged here. We duplicate Li's argument in favor of his thesis, and relax it to not condition on the Cobham-Edmonds thesis. If we relax feasibility from meaning exactly polytime computability, we explicate to some sort of bounded arithmetic, but its exact character remains imprecise. Assuming a different definition of feasible computation, Parikh's $I\Delta_0$ and Buss's $S_2^i, T_2^i$ theories are consistent with complementary versions of the postulates. Feasible mathematics may explicate to some formal system, but we are unable to say precisely which, if any, are the correct explicatum.

\item The relationship between $\PV$ and $\P$ is analogous to that between first-order logic and the general Turing machine. However, for the argument we presented previously, it is not immediately obvious if it is feasible, so we require a different argument. Cobham has provided a machine free characterization of the polytime computable functions, similar to that of primitive recursion. Cook defined the theory $\PV$ by scaffolding upon this definition so that function symbols may only be defined in $\PV$ according to this characterization. It is therefore easy to prove that $f$ is a function symbol of $\PV$ if and only if, in the standard model of $\PV$ that the interpretation of $f$ is polytime computable. 
\item We have presented this explication directly in section \ref{feasibleturingthesis} as the feasible version of Turing's thesis.
\end{enumerate}

Our argument is again limited in an identical situation as the feasible version of Turing's thesis and Gandy's thesis. We are unable to make $b'$ and $d'$ more precise than as presented.

\section{Failing to make the Explicatum Precise}
\label{complexity}

We have presented several kinds of arguments to support the Cobham-Edmonds thesis. First, in section \ref{prior}, we presented the standard arguments that appear in the literature. Note that none of these arguments imply that $\P$ is the correct explicatum, rather they are a body of empirical evidence to suggest that $\P$ as an explicatum is a plausible conclusion.

In section \ref{provingthethesis}, we showed that several explications in support of the Church-Turing thesis could be cloned and modified to serve as explications of the Cobham-Edmonds thesis, but these are all identically flawed. While Turing is able to explicate human computation to the Turing machine, we are able to explicate feasible computation only to a time-bounded Turing machine, and we are unable to justify the bound must necessarily be a polynomial. Turing, Gandy, and Kripke do not have this issue since it is obvious and convincing that their explicanda under consideration is as general as possible.

Our resolution is then this. We can assert that it is possible that $\P$ is the explicatum of feasible computation, but we cannot hope to assert that the explicatum of feasible computation must necessarily be $\P$. There are many candidates of more general classes that we can neither eliminate nor consider. We argue that the explications presented in section \ref{provingthethesis} must fail in principle.

\subsection{Why is Complexity Theory so Hard?}
The history of computational complexity theory is a history of failure, and rising from that failure. The $\P$ vs $\NP$ problem has eluded proof for half a century. There are many hard problems in mathematics: the Riemann hypothesis, Navier-Stokes, and so on. The $\P$ vs $\NP$ problem has a different mythology than these problems in that although complexity theorists are not able to prove it, they can prove broad classes of techniques cannot prove it. For example, in the early history of computational complexity, many proof techniques were just those used in computability theory, such as diagonalization. Diagonalization is a relativizing technique, and it was proven in \cite{baker1975relativizations} that relativizing techniques cannot resolve the $\P$ vs $\NP$ question. It was very surprising that the techniques that separated the computable from the computably enumerable languages were fundamentally incapable from separating their polynomially bounded counterparts. Rising from this barrier, complexity theorists developed combinatorial methods on circuits. If one could prove that $\NP$ did not have polynomial sized circuits, then $\P \neq \NP$. This program continued until the natural proofs barrier \cite{Razborov1997}. Most circuit techniques at the time were natural, and it was proven that the $\P$ vs $\NP$ problem cannot have a natural proof. This cycle of death and rebirth has repeated many times and in many more precise and specific ways. Specialists develop techniques to reach for the $\P$ vs $\NP$ problem. The techniques they develop are proven to have limitations in what kinds of problems they may apply to. From their failure arise new specialists with new techniques who promise not to make the mistakes of their ancestry.

There are several candidate classes $C$ such that the explication could reasonably explicate to $C$ with $\P \subseteq C$. For each such class, we argue that any argument, if made formal enough, would settle open complexity questions. The $\P$ vs $\NP$ problem is not really one problem, but several dozen problems connected through a vast web of implication. A solution to any could be significant. If such questions are provably hard, why should we expect there to be a routine explanation that anyone can provide? Barriers give the open problems in computational complexity a kind of limited independence, more so socially among specialists than as a formal notion of independence from a logical system. We are not stating that improving the precision of these explications is impossible. The following arguments are similar to that of a reduction. The hardness of improving these explications is inherited from the hardness of open complexity questions. Historic progress must be made elsewhere before it can be made here.

\subsection{Feasibility and Randomness}
\subsubsection{Randomized Computation}

A randomized Turing machine is a Turing machine with an extra tape that has been preloaded with an infinite sequence of uniformly random bits. These devices are used in the study of randomized algorithms. They are designed to separate the questions of computing with random bits and generating the random bits themselves. Define $\mathsf{BPTIME}[f(n)]$ (bounded error probabilistic $O(f(n))$-time) as the class of languages where there exists a randomized Turing machine $M$ deciding $L$ which halts in a $O(f(n))$ steps such that

\begin{align*}
w \in L &\iff P[M(w)\text{ accepts}] \geq 2/3 \\
w \not \in L &\iff P[M(w)\text{ accepts}] \leq 1/3
\end{align*}

Without this allowance of error, a perfect randomized algorithm could be made deterministic by replacing its randomness with some deterministic string of all zeroes and the answer should not change, trivializing the subject of study. Define $\BPP = \cup_{i=0}^\infty \mathsf{BPTIME}[n^k]$ as the class of languages decidable by polynomial-time randomized algorithms with bounded error. The threshold of $2/3$ is arbitrary since running $M$ a constant number of times and taking the majority answer will amplify this arbitrarily close to 1. Certainly $\P \subseteq \BPP$, but the question of whether $\P \neq \BPP$ or $\P = \BPP$ remains unsolved. The history of the $\P$ vs $\BPP$ problem is interesting in contrast to the $\P$ vs $\NP$ or $\P$ vs $\PSPACE$ problems. $\BPP$ was first defined by Gill \cite{gill1974computational}. It was later proven that $\BPP^A = \P^A$ for almost all oracles $A$, confirming that $\BPP$ is the correct definition of randomized computation \cite{bennett1981relative}. Gill was also the first to conjecture that $\P \neq \BPP$. He provided a general method for derandomization in exponential time, and did not think it could be significantly improved. Simultaneously, there was also a growing body of evidence that $\P \neq \BPP$. Several polynomial-time randomized algorithms were discovered for problems with no known deterministic polynomial-time algorithms, such as primality testing. The primality testing problem asks on input a number $n$, determine if $n$ is prime or composite. Miller was the first to give a deterministic polynomial-time primality testing algorithm, but its correctness was conditional on the unproven extended Riemann hypothesis \cite{miller1975riemann}. Rabin was able to remove the condition on the Riemann hypothesis by making the algorithm randomized \cite{rabin1980probabilistic}. In some sense, the complexity theoretic study of randomized algorithms began in order to study this premise replacement \cite{Wigderson2019}. There are numerous other examples of problems with polynomial-time randomized algorithms, and no known deterministic polynomial-time counterparts. Such evidence led a majority of complexity theorists to believe that $\P \neq \BPP$. 

Then came the hardness versus randomness paradigm. Impagliazzo and Wigderson proved that if $\TIME[2^{O(n)}]$ requires circuits of exponential size, then $\P = \BPP$ \cite{impagliazzo1997p}. General derandomization exists if you can prove that exponential time (with linear exponent) requires circuits of exponential size. Because the circuit lower bound is far more believable than the $\P \neq \BPP$, today an overwhelming number of computer scientists believe that $\P = \BPP$, but do so through clenched teeth. 

Finally, many strong examples of problems with polynomial time randomized algorithms and no known polynomial time deterministic algorithms fell. Agrawal, Kayal, and Saxena gave a fully deterministic polynomial-time algorithm for primality testing \cite{agrawal2004primes}. Their algorithm, ironically, came from the study of derandomization.

\subsubsection{The Explication of Feasible Randomized Computation}
How should the consideration of randomness affect any of our explications? Consider now a randomized version of the feasible version of Turing's thesis we presented in section \ref{feasibleturingthesis}. Our explicandum is a feasibly limited human computor with the addition of a coin they may flip only a feasible number of times. They may use the outcome of this coin flip to make decisions, and perhaps we allow them to make some mistakes. If we follow the feasible version of Turing's argument closely, we explicate onto some randomized complexity class, hopefully $\BPP$---but maybe $\mathsf{RP}$, $\mathsf{ZPP}$, or $\mathsf{PP}$---we cannot say precisely. 

We now argue that the feasible version of Turing's thesis cannot be made precisely onto $\P$. Suppose for the sake of argument that there exists a rigorous explication which demonstrates that the explicatum of feasible computation is necessarily $\P$. We suppose that such an explication is as convincing in the correctness of the explicatum, $\P$, as Turing's thesis was for the Turing machine. It must also be the case that what makes this argument necessarily rigorous will also explicate randomized feasible computation onto $\BPP$. Following Turing's argument, one can imagine the feasibly bounded human computor with a coin and distill and formalize it until one is left with a $\BPP$ machine. There are now two explications: feasible computation onto $\P$, and randomized feasible computation onto $\BPP$. 

For these two explications, we argue that the explicanda are distinct if and only if the explicata are distinct. First suppose the explicanda are distinct. If randomized feasible computation is unequal to feasible computation, there is some problem which has a feasible procedure whose feasibility requires the coin and what it entails, and this problem may not be feasibly computed without it. When transforming this intuitive procedure into a formal algorithm, one receives a $\BPP$ machine to compute this procedure. The problem this procedure solves cannot be in $\P$. Otherwise, a feasibly bounded human computor could simulate that $\P$-machine feasibly, contradicting our assumption that the feasibly bounded human computor cannot solve this problem. Since this problem is in $\BPP$ but not in $\P$, we observe that $\P \neq \BPP$. Conversely, if $\P \neq \BPP$, some problem is solvable by a $\BPP$ machine but not by a $\P$ machine. The feasibly bounded human computor with a coin may feasibly simulate this $\BPP$ machine, but the feasibly bounded human computor without the coin may not; otherwise, one could simply explicate this back onto a $\P$ machine. Thus the explicanda are distinct. 

How different are these two explications? Whatever precision we presume exists in the explication to rigorously conclude with $\P$ must have sufficient explanatory power to answer whether or not feasible algorithms are afforded randomness. However, this would immediately distinguish or equate the two explicanda and thus resolve the $\P$ vs $\BPP$ question. If the argument may deduce that randomized computation is necessarily infeasible, and the explication of feasible computation proceeds onto $\P$, then $\P \neq \BPP$. Similarly, if randomization gives no power to feasible computation, then the explication must also be able to conclude that $\P = \BPP$. This argument does not rely on the polynomial bounds themselves, as there are open problems in finer settings. If the explication is able to distinguish randomized computation from deterministic computation, but cannot necessarily argue that randomization provides a superpolynomial speedup, even proving $\TIME[n^2] \subsetneq \mathsf{BPTIME}[n^2]$ would be extremely significant. If the argument cannot necessarily say $\BPP \subseteq \P$, but can improve the best-known derandomization from $\BPP \subseteq \TIME[2^{poly(n)}]$ to $ \BPP \subseteq \TIME[2^{o(n)}]$, this would also be extremely significant.

In the general case of the Church-Turing thesis, it is not that we have several competing explications only because the explicata (Turing machine variants, the $\lambda$-calculus, and so on) can all be proven equivalent, and the explicanda all arguably stem from the same source. In the feasible case, the landscape is very different. We have several competing explications with possibly different conclusions. We can no more resolve that feasible computation must explicate onto $\P$ than we can conclude it must explicate onto $\BPP$, because we cannot yet solve the $\P$ vs $\BPP$ problem. Here, we have two similar, but possibly distinct explications, from two possibly distinct explicanda onto two possibly distinct explicata.
We would hope that the explication of feasible computation would have proceeded as rigorously and convincingly as the explication of computation in Turing's thesis; however, we can no more objectively assert that a single explication is the one true representative of feasible computation than we can prove $\P \neq \BPP$. It is not the case that the $\P$ vs $\BPP$ problem is unsolvable; rather, we do not expect a routine philosophical study to solve a highly technical and well-studied unsolved mathematical problem. 

Analogous arguments may be made for the role of randomness in Gandy's thesis. The original formulation by Gandy is entirely deterministic. In the feasible case, there is no obvious philosophical justification for us to include or exclude a machine whose local transitions are permitted to sample from some primitive source of physical randomness. Define the randomized feasible Gandy machine as a feasibly bounded Gandy machine with access to some physical randomness, which it may feasibly use in its computation. Suppose that the feasible version of Gandy's thesis could be strengthened with additional arguments so that the explicatum was necessarily $\P$. Then our randomized feasible Gandy machine could be equivalently explicated onto $\BPP$. Whatever additional justification existed to allow us to fully deduce the necessity of $\P$ must have sufficient explanatory power to be able to accept or deny the role of randomness in the computation of a feasible mechanism, but this would immediately settle the $\P$ vs $\BPP$ problem.

For the feasible version of Kripke's thesis, there are suitable theories of bounded arithmetic which aim to capture probabilistic feasible reasoning, including Je{\v{r}}{\'{a}}bek's $\APC_1$ \cite{Jerabek2007} or the more recently introduced $\mathsf{APX}_1$ \cite{10.1145/3798129.3800842}, but we focus on $\APC_1$ for simplicity. While $\PV$ is an equational theory, $\PV_1$ denotes the enlarged first-order theory to contain quantifier-free logical formulae and propositional reasoning \cite{Cook1975}. The theories $\PV$ and $\PV_1$ are often used interchangeably \cite{buss1985bounded}. Define the theory $\APC_1$ to be $\PV_1$ adjoined with the sentences $\mathsf{dWPHP}(f)$ for each $f$ in the language of $\PV$. The dual weak pigeonhole principle $\mathsf{dWPHP}(f)$ can informally be described as asserting that $f$ cannot be surjective if its co-domain is larger than its domain \cite{Oliveira2025}. This theory has been known to capture many lower bound results which make use of approximate counting methods, and are not known to be provable from $\PV_1$.

Suppose we were able to strengthen our feasible version of Kripke's thesis with additional justification, and we correctly explicated feasible mathematics onto $\PV$. It is then the case that whatever makes this argument convincing can also explicate probabilistic feasible reasoning onto $\APC_1$. Whatever rigorous understanding there is to perform this explication must be suitable to judge whether or not $\mathsf{dWPHP}(\PV)$ is either admissible into or banished from the intuitive notion of feasible mathematics, but this would obviously determine the relationship between the theories $\PV_1$ and $\APC_1$. Exactly how $\APC_1$ is conservative over $\PV_1$ (if at all) is a major unsolved and difficult problem in bounded arithmetic, and can be understood as a proof-theoretic version of the $\P$ vs $\BPP$ problem, although one does not necessarily entail the other. 

Although we have gone into great detail on the role of randomness and feasibilism, the root of our imprecision in our three explications is not the fault of randomness itself. The imprecise relationship between randomness and feasibilism is one of several questions. Every open structural complexity problem represents an analogous question.

\subsection{The Extended Church-Turing Thesis}
\label{ectt2}
Recall the statement of the extended Church-Turing thesis and our discussion from section \ref{ectt}. If a model of computation is considered ``reasonable'', then it may be simulated  by other reasonable models of computation with at most a polynomial overhead in the runtime of the simulator. There are many diverse kinds of machines but the intention of only characterizing some machines as reasonable is because these aught to model realistic computations. A machine model should be considered reasonable if the number of steps it takes is polynomially related to the real time an actualized physical machine may take to compute the same task. With the simplest Turing machine presumed reasonable, this defines a large class of related machines. In this section, we discuss the difficulties of making explicit what models of computation do or do not pass the ``reasonableness criterion''.

Consider the non-deterministic Turing machine with a polynomial-time bound. If an $\NP$ machine takes $T(n)$ steps, the best known simulation of this on a deterministic Turing machine takes $2^{O(T(n))}$ steps. If there was a convincing argument that the non-deterministic Turing machine should satisfy the reasonableness criterion, this would immediately imply $\NP \subseteq \P$. In fact, some incorrect attempts to prove $\P=\NP$ argue the physical realizability of $\NP$ computations, see \cite{Aaronson2005}. In the opposite intention, if one could say objectively that the $\NP$ machine is not reasonable, then this must prove $\P \neq \NP$. Determining if the $\NP$ machine satisfies the reasonableness criterion is as hard as the $\P$ vs $\NP$ problem. Improving the deterministic simulation of a non-deterministic machine, even to $2^{o(T(n))}$ would have drastic consequences. This would imply $\NP \neq \EXP$, $\EXP \not \subset \SIZE[poly]$, and several other open hard problems. 

Randomized computation is in a similar situation. If a $\BPP$ machine takes $T(n)$ steps, the best known simulation of this on a deterministic Turing machine is $2^{O(poly(T(n)))}$ steps \cite{gill1974computational}. If one could explicitly say the $\BPP$ machine does or does not satisfy the reasonableness criterion, this would immediately settle the open problem of $\P$ vs $\BPP$. 

\subsubsection{Quantum Computation}
We avoid technical discussion of quantum computers. Define $\BQP$ as the class of languages decidable by uniformly generated quantum circuits of polynomial size, with bounded error. The definition is set up such that $\P \subseteq \BPP \subseteq \BQP$. Quantum computers were first proposed as a fully theoretical model of computation. Interest grew significantly after Peter Shor proved that a quantum computer could solve the integer factorization problem in polynomial time \cite{shor1999polynomial}. If $p,q$ are distinct primes, the integer factorization problem asks on input $n = pq$ to output $p$ or $q$. Factorization has no known polynomial-time algorithm on a classical computer. In fact, RSA and other cryptographic protocols are secure only on the assumption that factorization is infeasible.   

With respect to the extended Church-Turing thesis, $\BQP$ is in an analogous situation as that of $\BPP$ and $\NP$. If a $\BQP$ machine has complexity $T(n)$, the best known simulation of this on a deterministic Turing machine takes $2^{O(poly(T(n)))}$ steps. Determining whether or not a $\BQP$ machine satisfies the reasonableness criterion would settle the $\P$ vs $\BQP$ question. This is as hard as any other question in structural complexity theory since $\P \subseteq \BPP \subseteq \BQP \subseteq \PSPACE$. If $\P = \BQP$, then $\P = \BPP$ and if $\P \neq \BQP$ then $\P \neq \PSPACE$. 

The increasing likelihood of an actualized quantum computer caused the extended Church-Turing thesis to come under closer scrutiny \cite{yao2003classical}. Several seemingly contradictory facts emerged simultaneously. Quantum computers can solve certain problems which are thought to be infeasible to ``reasonable" machines, so if a quantum computer could be actualized, is the reasonable criterion imprecise? There is a large diversity of opinion in this area. A Manhattan project level of funding has gone towards several teams of specialists attempting to actualize a quantum computer. Simultaneously, there are some prominent computer scientists who are critical of the actualizability of quantum computers, such as Oded Goldreich, Leonid Levin, and Gil Kalai. Theorems involving quantum computation stand true on their own, there is no disagreement there. Most criticism is derived from interpretations of physics, but Levin and Kalai have unique additional views relative to the extended Church-Turing thesis. Since quantum computers have a super polynomial speed up over machines the extended Church-Turing thesis characterizes as ``reasonable'', Levin takes the extended Church-Turing thesis as a system of belief, to mean that a quantum computer cannot be actualized \cite{Levin2003-LEVPTA-3}. Kalai believes the converse, that the quantum computer cannot be actualized affirms the extended Church-Turing thesis \cite{kalai2020argument}. It is possible that there might be a consistent theory to explain what is empirically observed about quantum computation, and simultaneously not violate the extended Church-Turing thesis \cite{aharonov2013quantum}. 

Since the reasonableness criterion cannot be made as precise as we would like, the extended Church-Turing thesis cannot be used as an independent source of feasibilism to derive the Cobham-Edmonds thesis from; the reverse is actually true. If one assumes the Cobham-Edmonds thesis, a reasonableness criterion may be easily defined: A model of computation is said to be reasonable if and only if it has been proven it may be simulated on a deterministic Turing machine with at most polynomial overhead. This is the practical working use of the extended Church-Turing thesis. Many machines types can be defined, and appear realistic on a first inspection, but if a machine model fails this version of the reasonableness criterion, they can be eliminated from consideration for the problem of study.

\section{Concluding Remarks}

\subsection{Open Problems}
The study of feasibilism has proceeded under numerous distinct names and regimes. Each paragraph in this section details a line of questioning that deserves further analysis than we were able to provide in this paper.

We have considered feasibility in an absolute sense; all problems must either be feasible or infeasible. This implicitly equates the notions of infeasibility and intractability. Although this most closely aligns with the use of feasibility, we should consider possible relaxations of this. First is with a tripartite demarcation: feasible, semi-feasible, intractable. Corresponding $\P$ with feasible and $\EXP$ with intractable---as most complexity theorists do---leaves algorithms with quasipolynomial run times in a state of uncertainty. Time complexities of the form $O(2^{\sqrt{n}})$ or $O(2^{(\log_2 n)^2})$ majorize every polynomial function and are majorized by every exponential function. The best-known algorithms for several important mathematical problems (including graph isomorphism \cite{Babai2016}) have quasi-polynomial run times. The class $\mathsf{QP} = \cup_{i=1}^\infty \TIME[2^{(\log_2 n)^i}]$ has been formally defined, but it remains unclear if this captures a notion which is distinguishable from intractability. Another possible relaxation is the consideration of feasibility as a measure. Problems are more or less feasible than others without any distinct classification of the feasible and infeasible. For example, Nguyen explicates feasibility as a fuzzy logical notion and assigns degrees of feasibility to algorithms by the growth rates of their time complexities \citep{nguyen1995algorithm}. He uses this to conclude that high degree polynomials are intuitively infeasible. 

McCulloch and Pitts construct and study the first theoretical model of a neuron. One of their lesser known contributions is the observation that if their model were affixed with a tape and the necessary machinery to fully access this tape, it would be equivalent to the Turing machine. They claim ``\textit{This is of interest as affording a psychological justification of the Turing definition of computability and its equivalents.}'' \cite{mcculloch1943logical}. Could the Cobham-Edmonds thesis be afforded such a psychological justification? Although feasibility appears more subjective than computability, perhaps there are some universal properties which may present themselves if a cognitive science or neuroscience approach is taken. 

For the Church-Turing thesis, there has been an effort to explore what unstated premises are necessary for the explications of Church, Turing, and others to succeed \cite{mundici1995paper, Sieg2008, dershowitz2008natural}. In the feasible case, our provided explications are incomplete. But were we to make further restrictions on the intuitive notions, these explications could certainly be completed. For example, in Turing's thesis, if we just assume the human computor is polynomial-time bounded, then obviously we may explicate onto a polynomial-time bounded Turing machine. The question is, what are the minimal additional assumptions we could make on the intuitive notions for our explications to succeed? Since we believe the explication must fail in principle, any such additional assumptions should be not too believable, or characterized by unsolved complexity conjectures. It is nonetheless an interesting question of what these could be. 

We considered feasibility as predicativity and as bounded constructivity, but there is a third trend that attempts to explicate feasibility in terms of theories of logic. In descriptive complexity, computational complexity classes are characterized by logical formalism; in particular, by the Immerman--Vardi theorem, polynomial-time over finite ordered structures is captured by first-order logic with least fixed points \citep{immerman1982relational,vardi1982complexity}. This does not by itself identify feasibility with $\P$, but it suggests a way of asking whether feasibility can be explicated through logical expressibility. Linear logic gives a related but proof-theoretic direction: refinements such as light linear logic were designed so that normalization/cut-elimination corresponds to polynomial-time computation \citep{girard1994light}. This serves as a possible framework for articulating feasibility through resource-sensitive logical structure. All of these represent potential directions to connect feasibility to logic directly.

The class $\P$ seeks to model problems with perfect feasible algorithms that are provably correct. There are many alternative working definitions of feasibilism with weaker requirements. Many instances of $\NP$-complete problems can be approximated efficiently by methods such as SMT solvers, simulated annealing, and gradient descent in neural networks. Such methods often perform well on structured or typical instances, but offer no worst-case guarantees, and require a different framework for analysis. Some problems have polynomial-time algorithms based upon unproven (and perhaps unprovable) but effective heuristics. It is also the case that algorithms can be considered feasible if the methodology of analysis is changed. The smoothed analysis model for example, can explain why although the simplex algorithm has a provable worst-case time complexity of $O(2^n)$, it terminates much earlier than this on practical instances \cite{10.1145/990308.990310}. It may be the case that a more robust treatment is possible by relaxation of what formal models are under consideration, and their requirements. 

Our diagrammatic presentation of Kripke's thesis in section \ref{kripke} allows for an additional observation. It is not only the case that the Church-Turing thesis can be deduced from assuming Hilbert's thesis, but that Hilbert's thesis can be deduced from Turing's thesis by the composition of $a + d + c$. This is notable, since Turing gives an extremely convincing argument of Turing's thesis. Although Hilbert's thesis is widely accepted, prior arguments in its favor have been criticized \cite{berk1982hilbert}. If combined with a proof that no major fragment of first-order logic is strong enough to encode the power of the Turing machine, this would suffice as a proof of Hilbert's thesis.

\subsection{Future Applications of this Analysis}
Our primary goal was to perform a detailed analysis of feasibilism, but we had a second goal. Feasibilism often serves as an unstated base upon which many other topics in theoretical computer science scaffold, and a strong analysis of feasibilism will assist a future analysis of those further topics. Consider the following quote from Avi Wigderson's book \textit{Mathematics and Computation}: 

\begin{quote}
I feel proud to belong to a field that has seriously taken on defining (sometimes redefining, sometimes in several ways) and understanding such fundamental notions that include: \textit{collusion, coordination, conflict, entropy, equilibrium, evolution, fairness, game, induction, intelligence, interaction, knowledge, language, learning, ontology, prediction, privacy, process, proof, secret, simultaneity, strategy, synchrony, randomness, and verification}. It is worthwhile reading this list again, slowly. I find it quite remarkable to contrast the long history, volumes of text written, and intellectual breadth that the concepts in this list represent, with the small size and the relative youthfulness of ToC, which has added so much to their understanding. -  \cite{Wigderson2019}
\end{quote}

We would like to emphasize two things. First, none of those words seemingly have anything to do with computers. Second, many of those topics take the Cobham-Edmonds thesis as a premise. With respect to any philosophical topic, there is always a question of whether or not a formal notion satisfies the criterion of material adequacy. Since theoretical computer scientists have to provide useful, working definitions, their definitions implicitly embody pragmatism as an implicit philosophy. Consider, for example, the definition of knowledge. Such a classical philosophical topic has been written and argued about for millennia. Ignoring all of this, cryptographers formulated their own definition of knowledge in order to build zero-knowledge proof systems \cite{goldreich1991proofs}. Knowledge, to a cryptographer, is a witness to an $\NP$-complete problem. Under the assumption that $\P \neq \NP$ and the Cobham-Edmonds thesis, this definition immediately provides immense explanatory power. Knowledge is infeasible (but not impossible) to obtain. Knowledge is feasible to transfer, in that one could feasibly prove to another that they have a certain piece of knowledge. These corollaries of the cryptographers' definition of knowledge require both a satisfactory definition of feasibility and the pragmatic philosophy they take in their working definitions. This is also not the only definition of knowledge in theoretical computer science. The field of distributed systems has an entirely distinct notion of common knowledge that is useful, and provides its own deep explanatory power \cite{halpern1990knowledge}. 

There are very rich subfields of theoretical computer science, each with its own definitions. Because these definitions are given value by their utility, they serve as highly convincing pragmatic explications of classical philosophical topics. We hope that this work convinces philosophers of science and analytic philosophers to more closely study the work being done in theoretical computer science, particularly computational complexity theory. 
\bibliographystyle{alpha}
\bibliography{references}
\end{document}